%% file: main.tex
\documentclass[journal]{ieeeaccess}

\usepackage{amssymb}
\usepackage{cite}
\usepackage{amsmath,amssymb,amsfonts}
\usepackage[pdftex]{graphicx}
\usepackage[bookmarks=false]{hyperref}
\usepackage{algorithmic}
\usepackage{algorithm}

\usepackage[utf8]{inputenc}

\usepackage{balance}
\usepackage{soul}
\usepackage{etoolbox}

\usepackage{ulem}

\usepackage{caption}
\usepackage{subcaption}

\graphicspath{{Images/}}

\DeclareGraphicsExtensions{.pdf}

\newtoggle{removeItalic}
\togglefalse{removeItalic} % regular command
\iftoggle{removeItalic} {
    \renewcommand{\textit}[1]{#1}
}

\def\BibTeX{{\rm B\kern-.05em{\sc i\kern-.025em b}\kern-.08em
    T\kern-.1667em\lower.7ex\hbox{E}\kern-.125emX}}

\begin{document}

\history{Received July 17, 2020, accepted August 11, 2020, date of publication August 17, 2020, date of current version August 28, 2020}
\doi{10.1109/ACCESS.2020.3017394}

\title{Cyberattacks on Miniature Brain Implants to Disrupt Spontaneous Neural Signaling}

\author{
    \uppercase{Sergio L\'opez Bernal}\authorrefmark{1},
    \uppercase{Alberto Huertas Celdr\'an}\authorrefmark{2},
    \uppercase{Lorenzo Fern\'andez Maim\'o}\authorrefmark{3},
    \uppercase{Michael Taynnan Barros}\authorrefmark{4},
    \uppercase{Sasitharan Balasubramaniam}\authorrefmark{5}, 
    \uppercase{and Gregorio Martínez Pérez}\authorrefmark{6}
}

\address[1]{Departamiento de Ingenier\'ia de la Informaci\'on y las Comunicaciones, University of Murcia, Murcia, Spain (e-mail: slopez@um.es)}
\address[2]{Telecommunication Software \& Systems Group, Waterford Institute of Technology, Waterford, Ireland (e-mail: ahuertas@tssg.org)}
\address[3]{Departamento de Ingenier\'ia y Tecnolog\'ia de Computadores, University of Murcia, Murcia, Spain (e-mail: lfmaimo@um.es)}
\address[4]{School of Computer Science and Electronic Engineering, University of Essex, UK and CBIG/BioMediTech in the Faculty of Medicine and Health Technology, Tampere University, Tampere, Finland (e-mail: michael.barros@tuni.fi)}
\address[5]{Telecommunication Software \& Systems Group, Waterford Institute of Technology, Waterford, Ireland (e-mail: sasib@tssg.org)}
\address[6]{Departamiento de Ingenier\'ia de la Informaci\'on y las Comunicaciones, University of Murcia, Murcia, Spain (e-mail: gregorio@um.es)}

% The paper headers
\markboth
{Author \headeretal: Preparation of Papers for IEEE TRANSACTIONS and JOURNALS}
{Author \headeretal: Preparation of Papers for IEEE TRANSACTIONS and JOURNALS}

\corresp{Corresponding author: Sergio L\'opez Bernal (e-mail: slopez@um.es).}

\begin{abstract}
Brain-Computer Interfaces (BCI) arose as systems that merge computing systems with the human brain to facilitate recording, stimulation, and inhibition of neural activity. Over the years, the development of BCI technologies has shifted towards miniaturization of devices that can be seamlessly embedded into the brain and can target single neuron or small population sensing and control. We present a motivating example highlighting vulnerabilities of two promising micron-scale BCI technologies, demonstrating the lack of security and privacy principles in existing solutions. This situation opens the door to a novel family of cyberattacks, called neuronal cyberattacks, affecting neuronal signaling. This paper defines the first two neural cyberattacks, Neuronal Flooding (FLO) and Neuronal Scanning (SCA), where each threat can affect the natural activity of neurons. This work implements these attacks in a neuronal simulator to determine their impact over the spontaneous neuronal behavior, defining three metrics: number of spikes, percentage of shifts, and dispersion of spikes. Several experiments demonstrate that both cyberattacks produce a reduction of spikes compared to spontaneous behavior, generating a rise in temporal shifts and a dispersion increase. Mainly, SCA presents a higher impact than FLO in the metrics focused on the number of spikes and dispersion, where FLO is slightly more damaging, considering the percentage of shifts. Nevertheless, the intrinsic behavior of each attack generates a differentiation on how they alter neuronal signaling. FLO is adequate to generate an immediate impact on the neuronal activity, whereas SCA presents higher effectiveness for damages to the neural signaling in the long-term.
\end{abstract}


\titlepgskip=-15pt

\maketitle

\input{tex/1introduction}
\input{tex/2related}
\input{tex/3technologies}
\input{tex/4attacks}
\input{tex/5useCase}
\input{tex/6results}
\input{tex/7conclusion}

\section*{Acknowledgement}

This work has been partially supported by the Irish Research Council, under the government of Ireland post-doc fellowship (grant code GOIPD/2018/466). We thank Luigi Petrucco, Ethan Tyler, and SciDraw for their publicly-available scientific images \cite{fig:mouse_brain, fig:mouse}.

\bibliographystyle{IEEEtran}
\bibliography{main}
\nocite{*}

\begin{IEEEbiography}[{\includegraphics[width=1in,height=1.25in,clip,keepaspectratio]{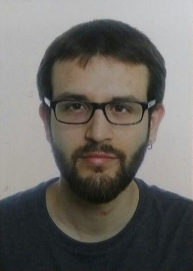}}]{Sergio L\'opez Bernal} is a PhD student at the University of Murcia. He holds a M.Sc. and a B.Sc. in Computer Engineering from the University of Murcia, and a M.Sc. in Architecture and Engineering for the IoT from IMT Atlantique, France. His research interests include ICT Security on Brain-Computer Interfaces, and Network and Information Security.
\end{IEEEbiography}

\begin{IEEEbiography}[{\includegraphics[width=1in,height=1.25in,clip,keepaspectratio]{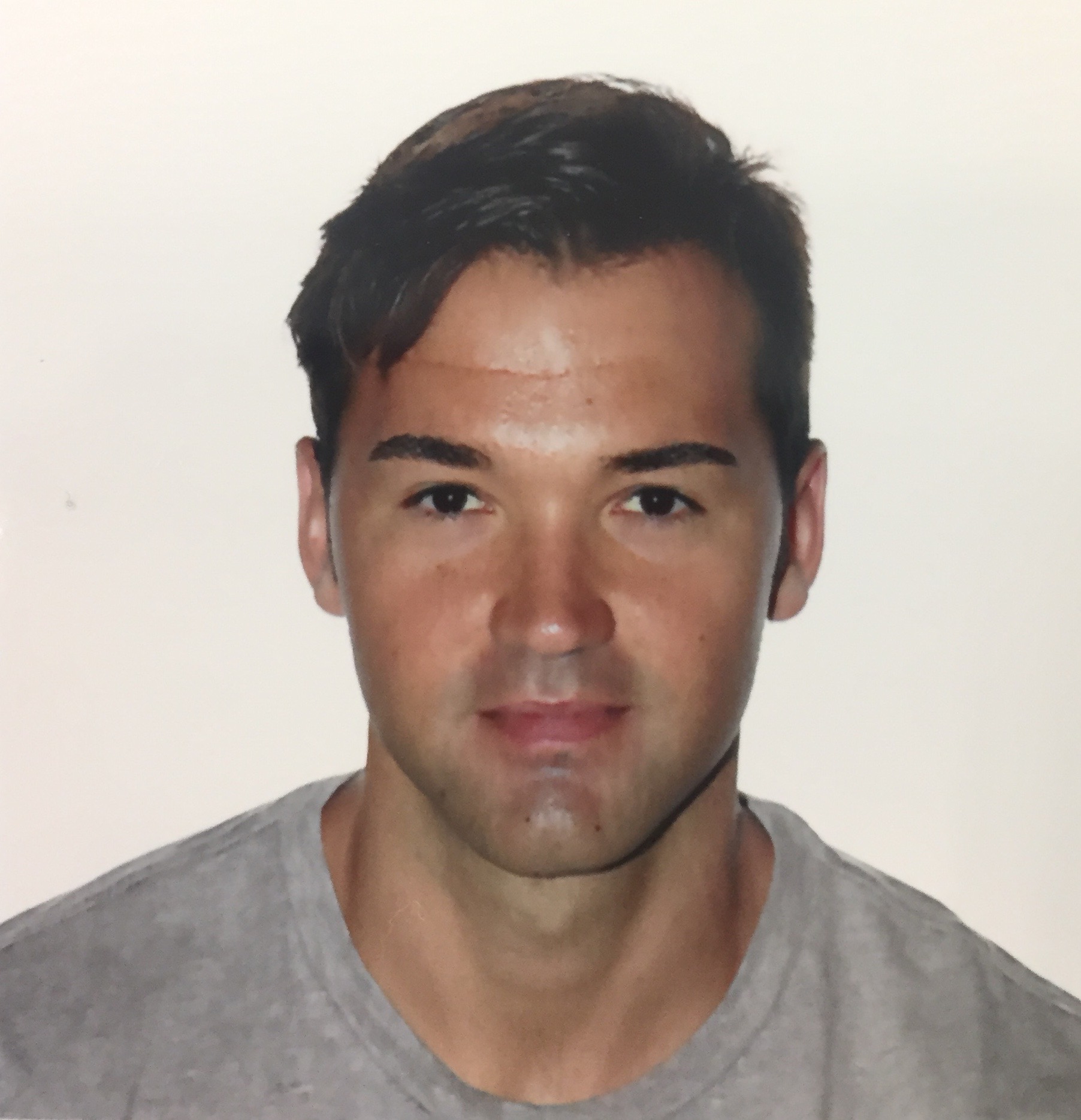}}]{Alberto Huertas Celdr\'an} is an Irish Research Council Government of Ireland Postdoctoral research fellow associated with the TSSG, Waterford Institute of Technology, Ireland. Huertas Celdr\'an received M.Sc. and Ph.D. degrees in Computer Science from the University of Murcia, Spain. His scientific interests include cybersecurity, privacy, Brain-Computer Interfaces (BCI), continuous authentication, and computer networks.
\end{IEEEbiography}

\begin{IEEEbiography}[{\includegraphics[width=1in,height=1.25in,clip,keepaspectratio]{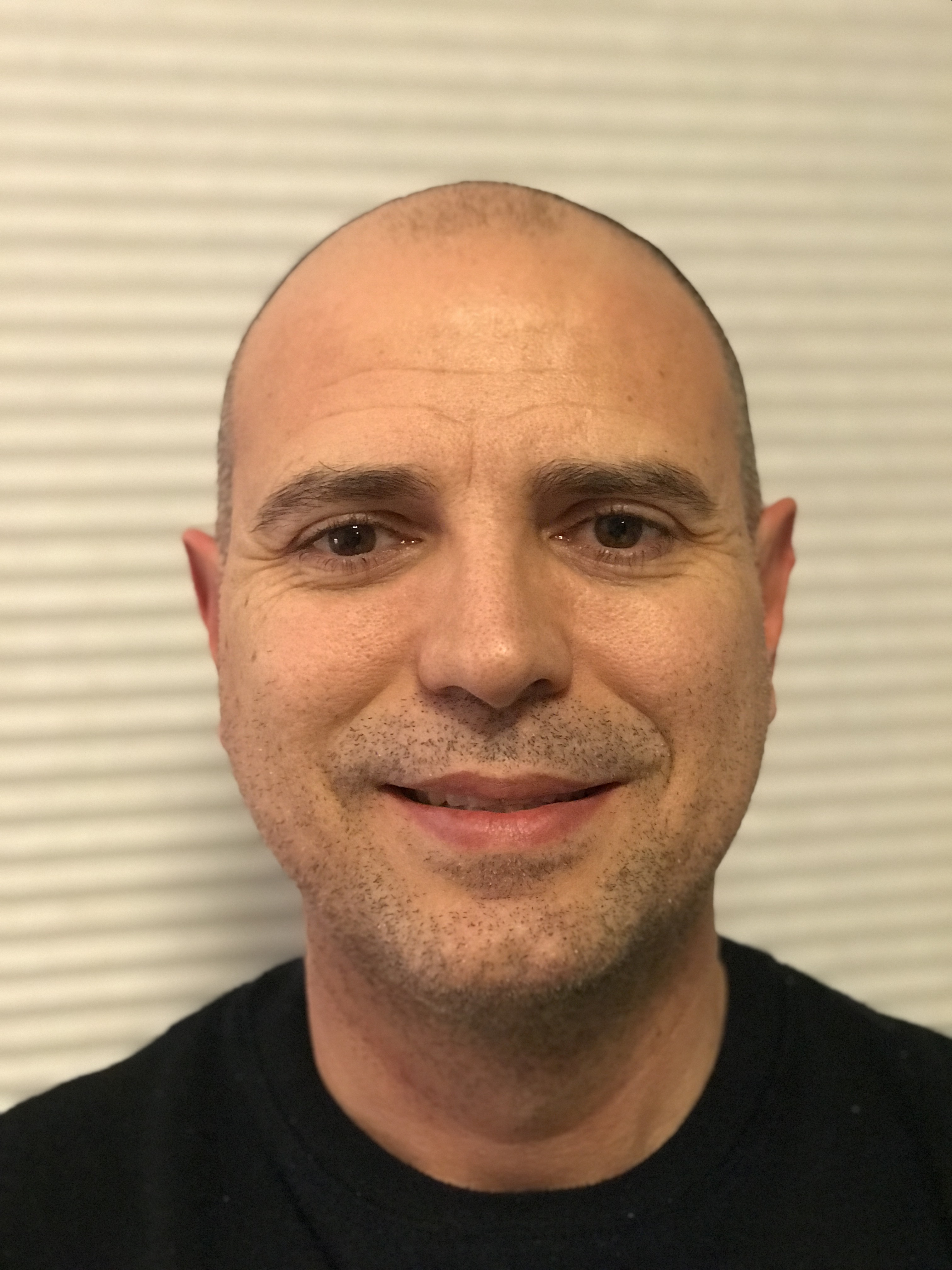}}]{Lorenzo Fern\'andez Maim\'o} is an associate professor in the Department of Computer Engineering of the University of Murcia. Fern\'andez Maim\'o has a MSc and PhD in Computer Science from the University of Murcia, Spain. His research interests primarily focus on machine learning and deep learning applied to cybersecurity and computer vision.
\end{IEEEbiography}

\begin{IEEEbiography}[{\includegraphics[width=1in,height=1.25in,clip,keepaspectratio]{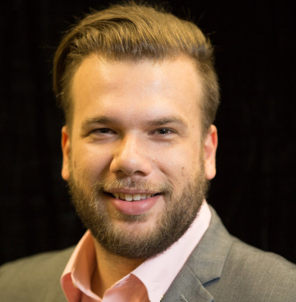}}]{Michael Taynnan Barros} is a Lecturer in the School of Computer Science and Electronic Engineering, University of Essex, UK and the recipient of the Marie Skłodowska Curie Individual Fellowship (MSCA-IF). He received his Ph.D. in Telecommunication Software at the WIT in 2016, M.Sc. degree in Computer Science at the Federal University of Campina Grande in 2012 and B.Tech. degree in Telematics at the Federal Institute of Education, Science and Technology of Paraiba in 2011.
\end{IEEEbiography}

\begin{IEEEbiography}[{\includegraphics[width=1in,height=1.25in,clip,keepaspectratio]{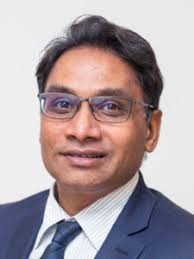}}]{Sasitharan Balasubramaniam} is the Director of Research with the TSSG, Waterford Institute of Technology, Ireland. He received the B.E. degree in electrical and electronic engineering from the University of Queensland, Brisbane, QLD, Australia, in 1998, the M.E. in computer and communication engineering from the Queensland University of Technology, Brisbane, in 1999, and the Ph.D. degree from the University of Queensland, in 2005. His current research interests includes molecular communications, Internet of (Bio) NanoThings, and Terahertz Wireless Communications.
\end{IEEEbiography}

\begin{IEEEbiography}[{\includegraphics[width=1in,height=1.25in,clip,keepaspectratio]{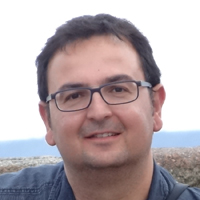}}]{Gregorio Mart\'inez P\'erez} is Full Professor in Department of Information and Communications Engineering of the University of Murcia, Spain. Mart\'inez P\'erez received M.Sc. and Ph.D. degrees in Computer Science from the University of Murcia, Spain. His scientific activity is devoted to cybersecurity, privacy, and networking, working on different national and European IST research projects on these topics.
\end{IEEEbiography}

\EOD

\end{document}

%% file: tex/1introduction.tex
\section{Introduction}
\label{sec:intro}

% 1- BCI definition
\IEEEPARstart{B}{rain-computer} Interfaces (BCIs) are considered as bidirectional communication systems between the brain and external computational devices. Although BCIs arose as systems focused on controlling external devices such as prosthetic limbs \cite{Lebedev:BrainMachineIF:2017}, they have gone one step further, enabling \textit{artificial} stimulation and inhibition of neuronal activity \cite{yao:stimulation:2019}. In the last years, neuronal stimulation has already been applied in different scenarios such as the provision of sensory feedback to prosthetic or robotic limbs \cite{doherty:nature:2011}, treatment of neurodegenerative diseases or disorders like Alzheimer's or depression \cite{kaufman:stimuli:2013}, and even futuristic applications such as interconnected networks of brains \cite{pais:2015:brainet} or brains connected to the Internet \cite{sempreboni:brainInternet:2018}. 

New BCI technologies are emerging, allowing a precise acquisition, stimulation, and inhibition of neuronal signaling. It reduces the brain damage caused by traditional invasive BCI systems and improves the limitations of non-invasive technologies such as attenuation, resolution, and distortion constraints \cite{ramadan:controlSignalsReview:2017,thomson:tsm:2019}. One of the most recent and promising BCI technique focuses on the use of nanodevices allocated across the brain cortex \cite{seo:neuralDust:2013}. Specifically, a relevant task of nanodevices equipped with optogenetic technology is the use of light to stimulate or inhibit engineered neurons according to different firing patterns sent by external transceivers \cite{Wirdatmadja:optogeneticModelProtocols:2017}. Promising initiatives such as Neuralink aim to accelerate the development of these technologies \cite{Musk2019}.

% 3- Cybersecurity problems and challenges
The previous BCI technologies hold the promise of changing our society by improving the cognitive, sensory, and communications skills of their users. However, they also open the door to critical cyberattacks affecting the subjects' safety and data security. In this context, essential vulnerabilities of current non-invasive BCI systems have been documented, exploited, and partially solved in the literature \cite{ballarin:cybersecurityAnalysis:2018}. As an example, the authors of \cite{Frank:subliminalProbe:2017, martinovic:feasibility:2012} demonstrated the feasibility of presenting malicious visual stimuli to extract subjects' sensitive data like thoughts. Besides, Sundararajan et al. \cite{Sundararajan:privacySecurityIssues:2017} conducted a successful jamming attack over the wireless communication used by the BCI, compromising its availability. However, the irruption of invasive and non-invasive stimulation and inhibition techniques, without security nor privacy capabilities, brings to the reality a novel family of cyberattacks affecting the neuronal activity. We call them \textit{Neural cyberattacks}, and they present a critical number of open challenges like the definition and categorization of the different neural cyberattacks and their neuronal behavior, the impact of each cyberattack to the neuronal behavior, and their consequences in the brain and body. \\

% 4- Contribution
To improve the previous challenges, the main contributions of this paper are the following ones:

\begin{itemize}
    \item The identification of cybersecurity vulnerabilities on emerging neurostimulation implants.
    \item To the best of our knowledge, the first description and implementation of neural cyberattacks focused on neuronal stimulation and affecting the activity of neural networks allocated in the human's brain. The proposed cyberattacks, \textit{Neuronal Flooding} and \textit{Neuronal Scanning}, are inspired by the behavior of current well-known cyberattacks in computer networks.
    \item The definition of three metrics to evaluate the impact of the two neural cyberattacks proposed: number of spikes, percentage of shifts, and dispersion of spikes.
    \item The implementation of the previous cyberattacks in a neuronal simulator to measure the impact produced by each one of them and the implications that they generate on the neuronal signaling. For that, we model a portion of a mouse's visual cortex based on the implementation of a CNN where the mouse is able to exit a maze.
\end{itemize}

% Structure of the paper
The paper remainder is organized as follows. Section~\ref{sec:relatedWork} gives an overview of the present state-of-the-art of current vulnerabilities, cyberattacks, and countermeasures affecting existing BCIs. After that, Section~\ref{sec:motivatingCase} illustrates emerging neurostimulation technologies and their cybersecurity concerns. Subsequently, Section~\ref{sec:definitionAttacks} offers a formal description of the cyberattacks proposed, while Section~\ref{sec:use_case} describes the implemented use case. Section~\ref{sec:results_analysis} first presents the metrics used to evaluate the impact of these cyberattacks, followed by the analysis of the results and impact that these cyberattacks generate. Finally, Section~\ref{sec:conclusion} briefly discusses the outcomes and potential future works.

%% file: tex/2related.tex
\section{Related work}
\label{sec:relatedWork}

% Cybersecurity on neurostimulation
During the last five years, new concepts such as brain-hacking, or neurocrime have emerged to describe relevant aspects of cybersecurity in BCI \cite{ienca:neuroprivacy:2015, ienca:hackingBrain:2016}. These works highlight that neuronal engineering devices, designed to stimulate targeted regions of the brain, would become a critical cybersecurity problem. In particular, they acknowledge that attackers may maliciously attempt to program the stimulation therapy, affecting the patient's safety. Furthermore, they emphasize that the cyberthreats do not need to be too sophisticated if they only want to cause harm. In this context, as indicated in this paper, it is possible to have a high impact on the brain by taking advantage of neurostimulation implants and send malicious electrical signals to the brain. Despite the identification of these risks, there are no studies in the literature defining or implementing neural cyberattacks, where the evaluation of their impact over the brain remains unexplored.
However, several vulnerabilities and attacks have been detected in BCI technologies performing neural data acquisition (e.g., EEG), which can serve as a starting point to perform neural cyberattacks. Section~\ref{sec:motivatingCase} offers additional considerations about vulnerabilities in BCI solutions.

% Known vulnerabilities and solutions
Platforms and frameworks that enable the development of BCI applications also present cybersecurity concerns, as demonstrated in \cite{takabi:privacyThreatsCounter:2016,bonaci:appStores:2015}. In this context, the authors of \cite{takabi:privacyThreatsCounter:2016} performed an analysis of the privacy concerns of BCI application stores, including Software Development Kits (SDKs), Application Programming Interfaces (APIs), and BCI applications. They discovered that most applications have unrestricted access to subjects' brainwave signals and can easily extract private information about their subjects. Moreover, Cody's Emokit project \cite{ienca:hackingBrain:2016}, managed to break the encryption of the Emotiv EPOC device (valid for all models before 2016), having access to all raw data transmitted. The authors of \cite{bonaci:appStores:2015} proposed a mechanism to prevent side-channel extraction of subjects' private data, based on the anonymization of neural signals before their storage and transmission.

% Existing attacks on neural data acquisition
The majority of the existing BCI systems are oriented to acquire, or record, neural data. Specifically, EEG BCI devices have gained popularity in recent years, due to their low cost and versatility, influencing the number of existing cyberattacks exploiting BCI vulnerabilities. In this context, the authors of \cite{li:bciApplications:2015} studied and analyzed well-known BCI applications and their potential cybersecurity and privacy concerns. Martinovic et al. \cite{martinovic:feasibility:2012} were able to extract users' sensitive information, such as debit cards or PINs, by presenting particular visual stimuli to the users and analyzing their P300 potential response. Another attack, performed by Frank et al. \cite{Frank:subliminalProbe:2017}, focused on presenting subliminal visual stimuli included within a video, aiming to affect the BCI users' privacy. Finally, in our previous work \cite{LopezBernal2019}, we studied the feasibility of performing cybersecurity attacks against the stages of the BCI cycle, considering different communication architectures, and highlighting their impact and possible countermeasures.

% Conclusion of the section
In conclusion, this section demonstrates that most of the related works are focused on presenting vulnerabilities and cyberattacks affecting the confidentiality, availability, and integrity of private data managed by BCIs. Nevertheless, there is a lack of solutions considering cyberattacks affecting the neuronal activity and, therefore, the subjects' safety. This article proposes two neural cyberattacks affecting the natural behavior of single and population of neurons.

%% file: tex/3technologies.tex
\section{Cybersecurity vulnerabilities of emerging neurostimulation implants}
\label{sec:motivatingCase}

This section introduces three promising BCI technologies capable of recording and stimulating neuronal activity with single-neuron resolution. For each scenario, we offer a description of its architecture, highlighting the cybersecurity vulnerabilities detected. Although these solutions are in an early stage, and they are still not commercial products, they are contemporary examples of how cybersecurity can affect existing and future implantable BCI solutions, and in particular for solutions that can target small neuron populations. These issues represent the starting point for the cyberattacks illustrated in the next sections of this paper. It is important to note that the objective of this section is not to find vulnerabilities in BCI devices or architectures but to justify how the proposed cyberattacks could be performed in realistic BCI systems. 

\subsection{Neuralink}
\begin{figure}[h]
\begin{center}
\includegraphics[width=\columnwidth]{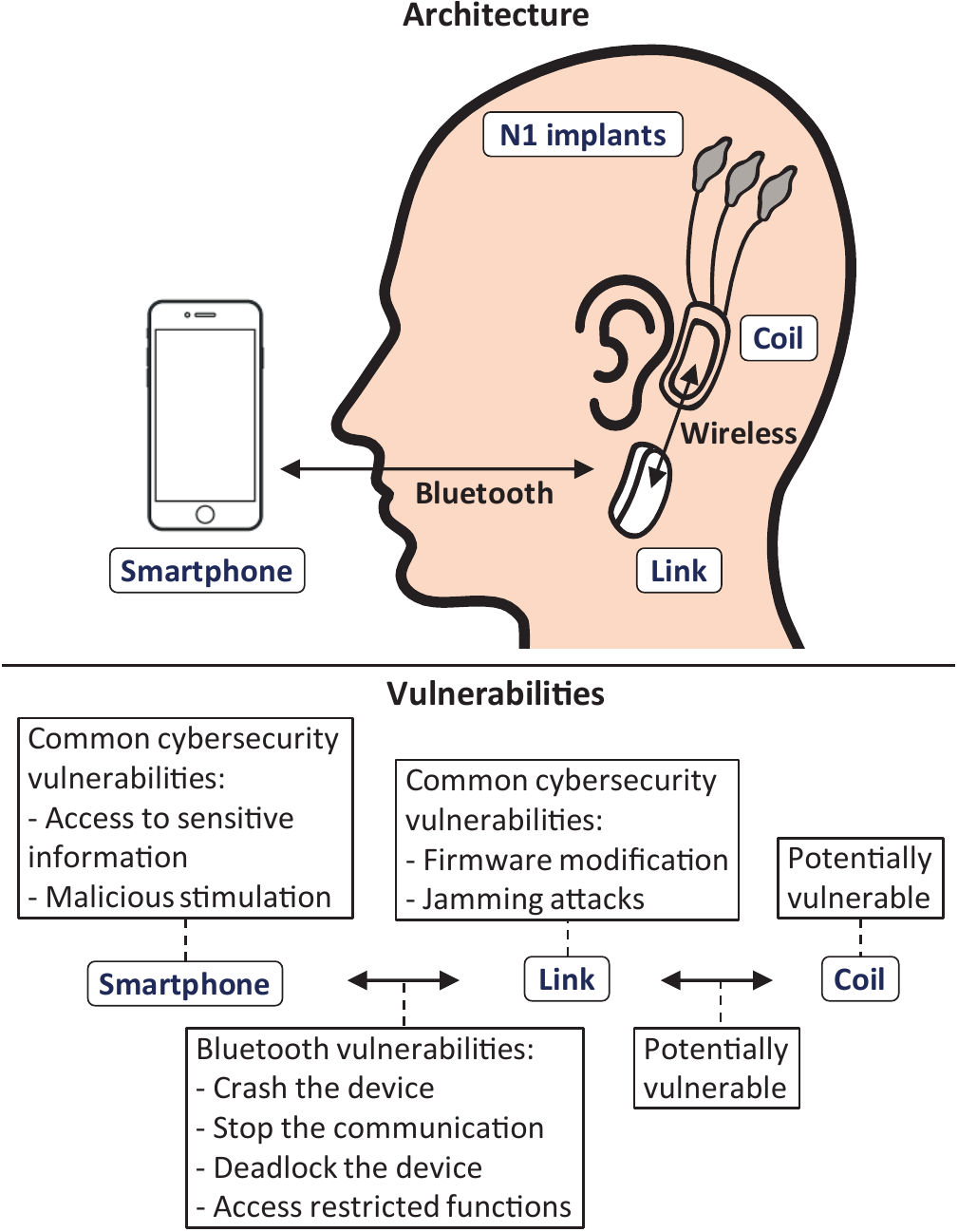}
\end{center}
\caption{Architecture and vulnerabilities of Neuralink.}
\label{fig:neuralink}
\end{figure}

Neuralink aims to record and stimulate the brain using new technologies, materials, and procedures to reduce the impact of implanting electrodes in the brain \cite{Musk2019}. The first element of the Neuralink architecture are the \textit{threads}, proposed as an alternative for traditional electrodes due to their biocompatibility, reduced size based on thin threads that are woven into the brain tissue, durability, and the number of electrodes per thread. Groups of threads connect to an \textit{N1 sensor}, a sealed device in charge of receiving the neural recordings from the threads and sending them stimulation impulses. With a simple medical procedure, up to ten N1 implants can be placed in the brain cortex. These devices connect, using tiny wires tunneled under the scalp, to a \textit{coil} implanted under the ear. The coil communicates wirelessly through the skin with a wearable device, or \textit{link}, placed under the ear. The \textit{link} contains a battery that represents the only power source in the architecture, deactivated if the user removes the link. \figurename~\ref{fig:neuralink} represents this architecture. 

% Cybersecurity
Although the communication mechanisms between the coil and the link are not provided, the link is managed via Bluetooth from external devices, such as smartphones, using an application. In this sense, Neuralink users can manage and personalize their links, upgrade their firmware, and include new security capabilities. We identify that this scenario can be potentially vulnerable as follows. First, the wireless mechanism used in the communication between the coil and the link could be vulnerable, depending on the protocol used \cite{Figueroa2019}. Besides, the Bluetooth communication between the smartphone and the link can also be vulnerable, according to the version used \cite{Sevier2019, Zubair2019}. As an example, we identify SweynTooth, a set of 12 vulnerabilities affecting a large number of devices using Bluetooth Low Energy (BLE) technologies. Based on them, an attacker could crash the device and stop its communications \cite{CVE:BLEcrash:2019}, deadlock the device \cite{CVE:BLEdeadlock:2019}, or access functions only available for authorized users \cite{CVE:BLEfunction:2019}. 

Moreover, the external device manages the logic of both acquisition and stimulation processes, including into these scenarios its inherent risks, and becoming one of the most sensitive elements of the architecture. In particular, Li et al. \cite{li:bciApplications:2015} detected that attackers could take total control of a smartphone running a BCI application, getting access to sensitive information, or performing malicious stimulation actions. Furthermore, the \textit{link} is a critical element of the architecture, where attackers can modify the firmware of the device to have a malicious behavior, as identified by \cite{pycroft:brainjacking:2016} for brain implants or to perform jamming attacks to disrupt the communication between devices, described by \cite{Vadlamani2016} for wireless networks. 

\subsection{Neural dust}

\begin{figure}[h]
\begin{center}
\includegraphics[width=\columnwidth]{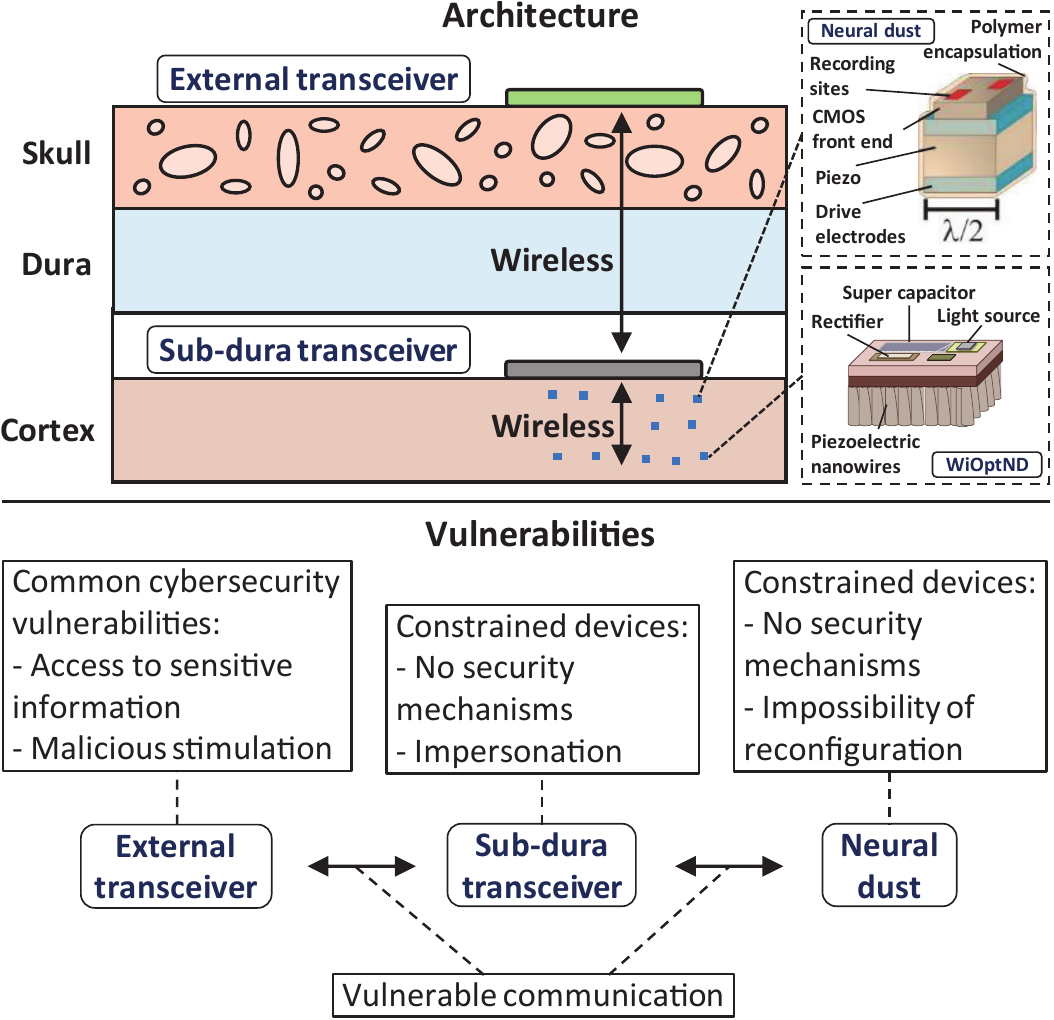}
\end{center}
\caption{Architecture and vulnerabilities of Neural dust.}
\label{fig:neural_dust}
\end{figure}

This architecture is composed of millions of resource-constrained nanoscale implantable devices, also known as neural dust, floating in the cortex, able to monitor neural electrophysiological activity precisely \cite{seo:neuralDust:2013}. These devices communicate with the sub-dura transceiver, a miniature device (constructed from components that are built from nanomaterials) placed beneath the skull and below the dura mater. This device uses two different transceivers to: (1) power and establish communication links with the neural dust, (2) communicate with external devices. During neural recording, the sub-dura transceiver performs both spatial and frequency discrimination with sufficient bandwidth to power and interrogate each neural dust. The external transceiver is a device without computational and storage restrictions, allocated outside of the patient's head. Wearables, smartphones, or PCs are examples of this device. The main task of the external transceiver is to power and communicate with the sub-dura transceiver and to receive the neuronal behavior from the sensing by the neural dust. \figurename~\ref{fig:neural_dust} presents the architecture of this solution, as well as the potential vulnerabilities that it presents.

Nevertheless, this technology has not been conceived following the principle of security and privacy by design. As a consequence, these devices do not implement authentication mechanisms to prevent malicious users from collecting neural sensing data from the neural dust, and they do not protect the transmitted data. In particular, the neural dust are resource-constrained devices without computational and storage capabilities to execute security functionalities like authentication protocols, ciphered communications, or data encryption. In this sense, external attackers could power and communicate to the implants to monitor private neural data. Finally, the sub-dura and external transceivers do not implement authentication protocols nor security mechanisms. An attacker could impersonate the external transceiver to communicate with the sub-dura device, and obtain sensitive neuronal signaling.

\subsection{Wireless Optogenetic Nanonetworks}

The Wireless Optogenetic Nanonetworking device (WiOptND) \cite{ Wirdatmadja:optogeneticModelProtocols:2017} is an extension from the neural dust \cite{seo:neuralDust:2013} but with the capability of optogenetically stimulating the neurons. Optogenetic stimulation uses light to stimulate neurons genetically engineered with specific genes that are sensitive to signals at a particular wavelength. This in turn provides targeted stimulation of very small population of neurons that have been engineered, enabling precise targeting of neural circuits within the micro-columns. Similar to the architecture of the neural dust, the WiOptND also receives power that is emitted from the sub-dura, which in turn communicates to the external transceiver. However, since the WiOptND is responsible for stimulating the neurons, the external transceiver will communicate the sequence of firing the neurons to the sub-dura transceiver to synchronize the charging and communication of the WiOptND implants. This is achieved by sending the firing sequence, in the form of a raster plot, to the external transceiver. This opens up new opportunities for attackers to send malicious firing patterns into the external transceiver, which will produce a new sequence of firing patterns for neural stimulation, resulting in detrimental consequences for the brain. Finally, the architecture and vulnerabilities described in \figurename~\ref{fig:neural_dust} also apply for WiOptND.

In conclusion, the previous vulnerabilities raise different concerns affecting the integrity, confidentiality and availability of subject's neural data. These vulnerabilities motivate different attack vectors to perform the neural cyberattacks described in subsequent sections.

%% file: tex/4attacks.tex
\section{Definition of neural cyberattacks}
\label{sec:definitionAttacks}

% Overview
Once demonstrated the feasibility of stimulating individual neurons by attacking different technological solutions, we formally describe two cyberattacks, \textit{Neuronal Scanning} and \textit{Neuronal Flooding}, aiming to maliciously affect the natural activity of neurons during neurostimulation procedures. They are inspired by the behavior and goals of some of the most well-known and dangerous cyberattacks affecting computer networks.

To formalize both cyberattacks, we denote $\mathbb{NE} \subset \mathbb{N}$ as a subset of neurons from the brain, where $n \in \mathbb{NE}$ expresses every single neuron. The voltage of a single neuron in a specific instant of time is denoted as $v_n\in\mathbb{R}$, whereas $vi_n\in\mathbb{R}$ indicates the voltage increase used to overstimulate a neuron $n$. Moreover, $\textbf{t}^{\text{win}}$ represents a temporal window in which the cyberattack is performed, equivalent to the duration of the simulation in Section~\ref{sec:results_analysis}. $\textbf{t}^{\text{attk}}$ is the time instant when the cyberattack starts, and $\Delta t$ the amount of time between evaluations during the process. In the implementation of the cyberattacks, it represents the duration of the steps of the simulation.

% Neuronal Flooding
\subsubsection{Neuronal Flooding}
\label{subsec:definitionFlooding}

In the cyberworld, a flooding cyberattack is designed to bring a network or service down by collapsing it with large amounts of network traffic. Traffic is usually generated by many attackers and forwarded to one or more victims. Extrapolating this network cyberthreat to the brain, a Neuronal Flooding (FLO) cyberattack consists in stimulating multiple neurons in a particular instant of time, changing the normal behavior of the stimulation process and generating an overstimulation impact. The execution of this cyberattack does not require prior knowledge of the status of the affected neurons since the attacker only has to decide what neurons to stimulate and when. This fact makes this cyberattack less complex than other cyberattacks that require prior knowledge of the neuronal behavior.

In particular, FLO performs the overstimulation action at $\textbf{t}^\text{attk}$. In that precise moment, a subset of neurons $\mathbb{AN} \subseteq \mathbb{NE}$ is attacked. This cyberattack is formally described in Algorithm \ref{alg:formalFLO}.

\begin{algorithm} 
\caption{FLO cyberattack execution} 
\label{alg:formalFLO} 
\begin{algorithmic}
    \STATE $t=0$
    \WHILE{$t < t^{win}$}
        \IF{$t == t^{attk}$}
            \FORALL{$n \in \mathbb{AN}$}
                \STATE $v_n \leftarrow v_n+vi_n$
            \ENDFOR
        \ENDIF
        \STATE $t \leftarrow t + \Delta t$
    \ENDWHILE
\end{algorithmic}
\end{algorithm}

\begin{figure}[h]
\begin{center}
\includegraphics[width=\columnwidth]{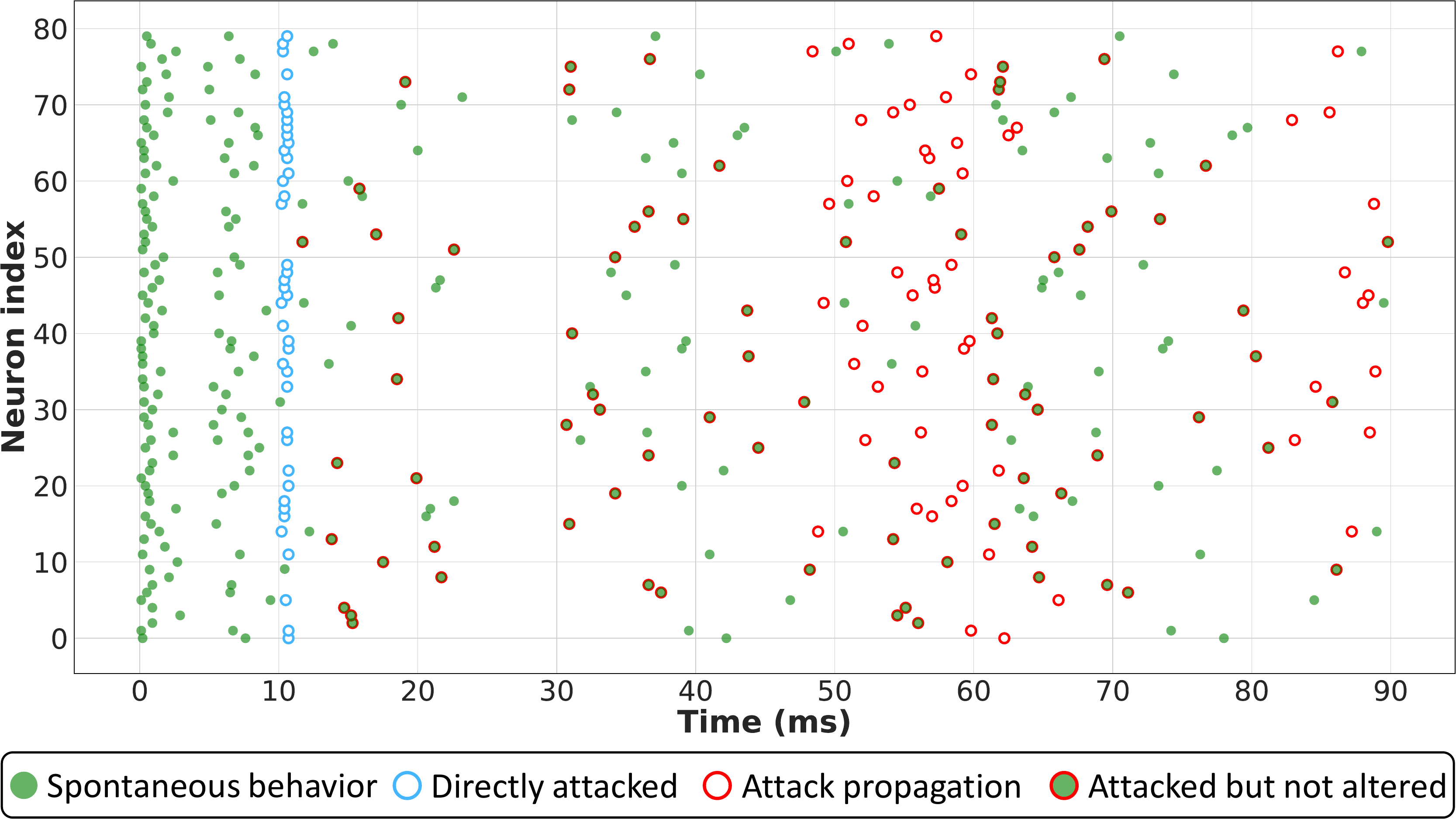}
\end{center}
\caption{Raster plot of a FLO cyberattack when the attack is performed at 10ms.}
\label{fig:raster_FLO}
\end{figure}

\figurename~\ref{fig:raster_FLO} represents an example to appreciate graphically the behavior of a FLO cyberattack, where the details of the neuronal network used in the simulation are not relevant at this point (addressed in Section~\ref{sec:use_case}). In particular, it represents the comparison of the FLO cyberattack with the spontaneous behavior for a simulation of 80 neurons, a duration of 90ms, and 42 neurons attacked in the instant 10ms. Green dots represent the neuronal spontaneous behavior, blue circles indicate the instant when the neurons are attacked, red circles highlight the propagation of the cyberattack in time, and those dots with a green color and red outline represent spikes common to both spontaneous and under attack situations. In this figure, we can see that all the attacked neurons alter their behavior, having spikes in different moments compared to the spontaneous activity.

% Neuronal Scanning
\subsubsection{Neuronal Scanning}
\label{subsec:definitionScanning}

Port scanning is another well-known cybersecurity technique performed by attackers to discover vulnerabilities in operating systems, programs, and protocols using network communications. In particular, it aims to test every networking port of a machine, checking if it is open and discovering the protocol or service available in that end-point. In the brain context, a Neuronal Scanning (SCA) cyberattack stimulates neurons sequentially, impacting only one neuron per instant of time. Based on that, it is essential to note that attackers do not require prior knowledge of the neuronal state to perform neural scanning cyberattacks. This fact, together with the stimulation of one neuron per instant of time, makes a low attack complexity.

Considering the notation previously defined, Algorithm \ref{alg:formalSCA} describes an SCA cyberattack. In particular, it sequentially overstimulates all the neurons included in the set of neurons $\mathbb{NE}$, without repetitions. For each neuron $n$, its voltage $v_n$ increases by $vi_n$. It is essential to indicate that the conditional clause limits the instants in which an attack can be performed, where $t^{attk}$ represents the attack over the first neuron of the set, and $t^{attk}+|NE|\Delta t$ the attack over the last neuron. 

\begin{algorithm} 
\caption{SCA cyberattack execution} 
\label{alg:formalSCA} 
\begin{algorithmic}
    \STATE $t=0$
    \WHILE{$t<t^{win}$}
        \IF{$t \in [t^{attk},t^{attk}+|NE|\Delta t]$}
            \STATE $n \leftarrow (t-t^{attk})/\Delta t$
            \STATE $v_n \leftarrow v_n+vi_n$
        \ENDIF
        \STATE $t \leftarrow t + \Delta t$
    \ENDWHILE
\end{algorithmic}
\end{algorithm}

\begin{figure}[h]
\begin{center}
\includegraphics[width=\columnwidth]{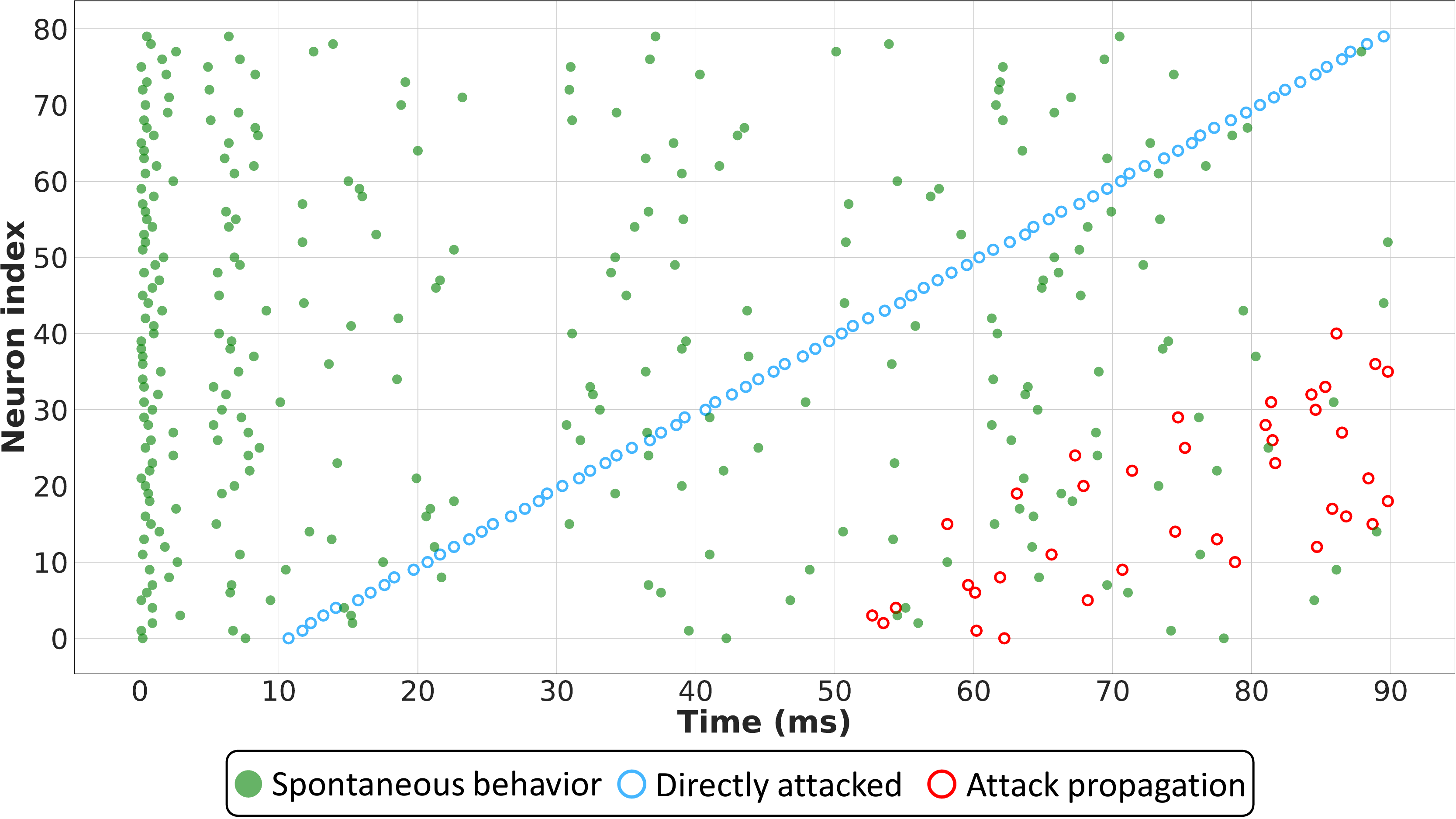}
\end{center}
\caption{Raster plot of an SCA cyberattack, from the instant 10ms to 90ms.}
\label{fig:raster_SCA}
\end{figure}

Finally, \figurename~\ref{fig:raster_SCA} shows, in a visual way, the behavior of an SCA cyberattack. We simulate 80 neurons during 90s, and sequentially attack all neurons, starting in the instant 10ms. The color code followed is the same as in \figurename~\ref{fig:raster_FLO}. As can be seen, the sequential attack of the neurons generates a diagonal line in the spikes. All spikes over the line remain unaltered since those neurons have not yet been affected by the attack. On the contrary, the spikes under the diagonal are affected by the attack. 

%% file: tex/5useCase.tex
\section{Exploiting vulnerabilities due to cyberattacks}
\label{sec:use_case}

This section introduces the use case used to implement the cyberattacks defined in Section \ref{sec:definitionAttacks}. We present the scenario and the experimental setup implemented to create the neuronal topology required to test the cyberattacks.

\subsection{Use case and experimental setup}
\label{subsec:experimentsDeployments}

The knowledge of precise neocortical synaptic connections in mammalian is nowadays an open challenge \cite{Gal2017}. Based on this absence of realistic neuronal topologies, we have studied the primary visual cortex of mice and replicated a portion of it, modeled using a Convolutional Neural Network (CNN)~\cite{Geron2019}. This CNN was trained by means of reinforcement learning \cite{Sutton2018} to represent a simple system able to make decisions based on a maze and find its exit. As indicated by Kuzovkin et al. \cite{Kuzovkin2018}, CNNs, and biological neuronal networks present certain similarities. First, lower layers of a CNN explain gamma-band signals from earlier visual areas, whereas higher layers explain later visual regions. Furthermore, early visual areas are mapped to convolutional layers, where the fully connected layers match the activity of higher visual areas. That is to say, the visual recognition process in both networks is incremental and move from simple to abstract. At this point, it is essential to note that we cannot compare the topology and functionality of a CNN to the complexity of the neuronal connections of a real brain. We only used this technique to provide a simple topology that is then implemented in a neuronal simulator to evaluate how attacks over a simplistic but realistic environment can affect the activity of simulated neurons, as indicated in Section~\ref{subsec:neuronalSimulation}.

\begin{figure}[h]
\begin{center}
\includegraphics[width=0.7\columnwidth]{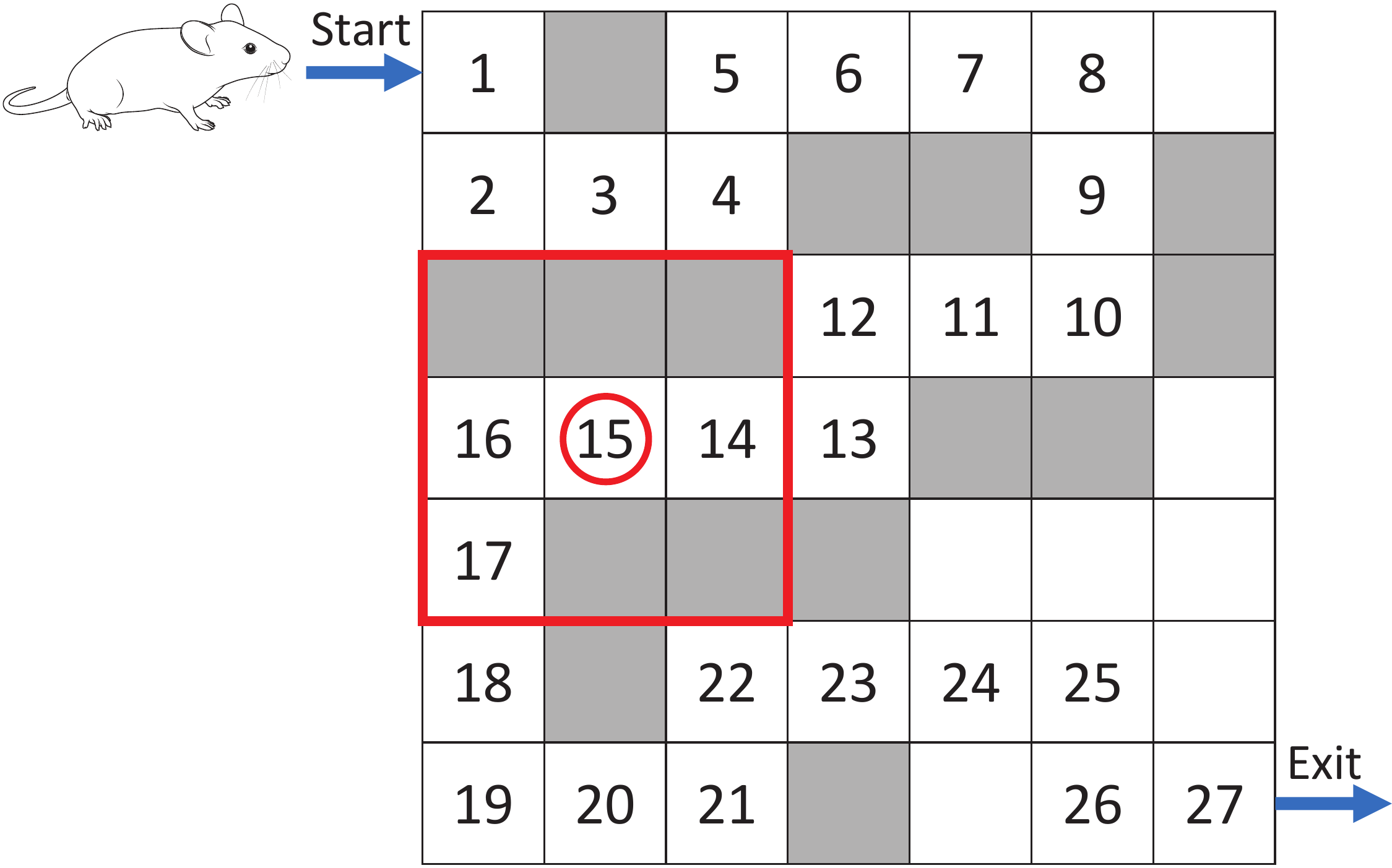}
\end{center}
    \caption{Maze used in our use case to model the movement of the mouse, including the optimal path between the starting and final cells. There are nine visible positions from the cell 15, highlighted within a red square.}
\label{fig:mazeLayout}
\end{figure}

In this context, we designed a simple proof of concept based on the idea of a mouse that has to solve the problem of finding the exit of a particular maze, inspired in the code from \cite{Zafrany}. The mouse must find the exit with the smallest number of movements and starting from any position. We define a maze of 7x7 coordinates, as represented in \figurename~\ref{fig:mazeLayout}. It contains one starting position identified with "1", while the exit is labeled with "27". Moreover, the positions colored in gray represent obstacles, and those in white are accessible positions through which the mouse can move. In this scenario, the mouse can move in all four 2D directions: up, down, left, and right. The numbering from 1 to 27 defines the optimal path determined by the trained CNN to reach the exit position, considering the lowest number of steps. Finally, it is essential to define the concept of \textit{visible position}. From each particular cell of the maze, the mouse can visualize a square of 3$\times$3 adjacent positions, including those that represent obstacles. This situation is highlighted in \figurename~\ref{fig:mazeLayout} with a red square, indicating the visible positions from the cell 15 of the optimal path.

\subsection{Convolutional Neural Network}
\label{subsec:CNN}

Our objective was to generate a CNN able to exit the maze from any position. We also aimed to define a topology with a reduced number of nodes to be compatible with resource-constrained neuronal simulators since we aim to evaluate this topology in multiple simulators. Nevertheless, for simplicity, this work includes  details of the implementation in only one simulator, as described in Section~\ref{subsec:neuronalSimulation}. To solve our maze problem, we implemented a CNN composed of two convolution layers and a dense layer. The ensemble of these three layers defines a complete CNN of 276 neurons, representing a small portion of a mouse primary visual cortex, summarized in Table~\ref{table:CNN}. We implemented this CNN using Keras on top of TensorFlow \cite{Keras2015}.

\begin{table}[h]
\caption{Summary of the layers of the CNN}
\label{table:CNN}
\setlength\tabcolsep{2pt}
\resizebox{\columnwidth}{!}{
\begin{tabular}{|c|c|c|c|c|c|c|c|c|}
\hline

Layer & Type & Filters & \begin{tabular}{@{}c@{}}Input \\ size\end{tabular} & \begin{tabular}{@{}c@{}}Output \\ size\end{tabular} & \begin{tabular}{@{}c@{}}Kernel \\ size\end{tabular} & Stride & \begin{tabular}{@{}c@{}}Activation \\ function\end{tabular} & Nodes \\ \hline

1 & Conv2D & 8  & 7$\times$7$\times$1 & 5$\times$5$\times$8 & 3$\times$3 & 1 & ReLU & 200 \\ \hline
2 & Conv2D & 8  & 5$\times$5$\times$8 & 3$\times$3$\times$8 & 3$\times$3 & 1 & ReLU & 72 \\ \hline
3 & Dense  & -  & 3$\times$3$\times$8 & 4 & - & - & ReLU & 4 \\ \hline

\end{tabular}}
\end{table}

\begin{figure}[h]
\begin{center}
\includegraphics[width=\columnwidth]{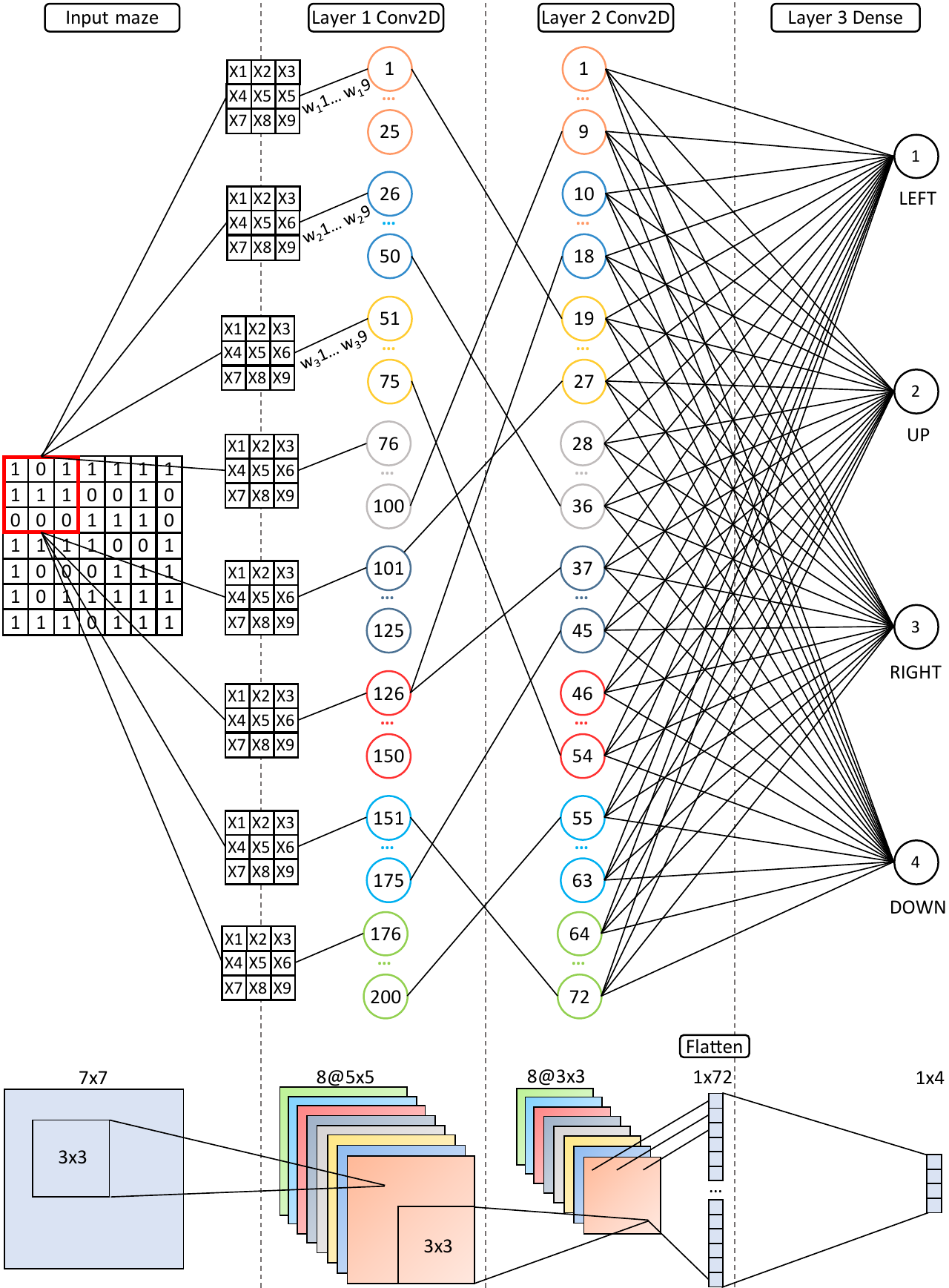}
\end{center}
\caption{Visual representation of the implemented CNN. It introduces a simplifications of the whole topology, indicating how the convolution process is performed and how nodes connect between layers. The color of each node matches the color of its associated filter.}
\label{fig:CNN_topology}
\end{figure}

\figurename~\ref{fig:CNN_topology} depicts the architecture of the implemented CNN which is also described in Table~\ref{table:CNN}. In particular, we have included a first 2D convolution layer with a $3\times 3$ kernel. This layer takes as input the current status of the maze, focusing each neuron on a square of 9 ($3\times3$) adjacent positions. In our experiments we determined that 8 filters of size $3\times3$ in each layer were sufficiently expressive. To represent the maze, each position contains a 1 value if the position is accessible, a 0 value if it is an obstacle, or a 0.5 value in the position of the mouse. 

During the training, each filter of the first layer specializes on a particular aspect of the maze. For example, a filter could focus on detecting vertical walls, while another could detect corners. The filters of the second layer can detect more complex scenarios by composing the output of these initial detectors. Since the input is a $7\times7$ maze, and the kernel is $3\times3$, the first convolution process requires 25 neurons ($5\times5$ kernel outputs) to cover the new $5\times5$ subset of the maze on the next layer. Since we use 8 different filters, the total number of neurons required to produce the first layer's output of the CNN is 200 ($5\times5\times8$). This is illustrated in \figurename~\ref{fig:CNN_topology}, where each group of neurons has a different color that matches the color of its filter. Therefore, since the first layer generates an output of size $5\times5\times8$, the application of the $3\times3$ kernels of the second convolutional layer requires a total of 72 ($3\times3\times8$) neurons. Finally, this new output is sent through a last dense layer of 4 neurons, one for each possible movement direction on the maze (left, up, right, down). Each output is an estimation of the probability of success with each movement, being selected the direction with the greatest score. 

In order to understand Section~\ref{subsec:neuronalSimulation} and Section~\ref{sec:results_analysis}, it is necessary to explain the mapping between the sequential number of each neuron and its position in its associated filter output. \figurename~\ref{fig:CNN_topology} shows this mapping. Each neuron have associated a 3-dimensional vector, where the third coordinate is its filter and the two first coordinates, the position in that filter output. The order is as follows: the first neuron has the coordinates [0,0,0], corresponding to the first neuron in the first filter output; the eighth neuron corresponds to [0,0,7]; the ninth one is [0,1,0], and so on until the 200th neuron, with coordinates [4,4,7].

\subsection{Biological neuronal simulation}
\label{subsec:neuronalSimulation}

After training the CNN, we represented its resulting topology in Brian2, a lightweight neuronal simulator \cite{Stimberg2019}. We selected Brian2 because it is adequate to run neuronal models in user-grade computers, without the requirement of using multiple machines, or even supercomputers. It also presents a good behavior in the implementation of neuron models with simplified and discontinuous dynamics (such as Leaky Integrate-and-Fire or Izhikevich) \cite{tikidji-hamburyan:simulators:2017}. Other alternatives, such as NEURON, present complex solutions to model neurons with fine granularity, offering distributed computation capabilities for high demanding simulations. Nevertheless, this functionality is unnecessary in our particular study.  

We maintain in the biological simulation the exact number of layers, the number of neurons per layer, and the topological connections between neurons. However, there is a crucial difference between the implementation of these two approaches. In the CNN, a filter weight represents the importance that a connection between two neurons of different layers have on the topology and, thus, over the solution. In the biological simulation, we transform the CNN weights to synaptic weights, representing the increase of the voltage induced during an action potential. Table~\ref{table:comparison_CNN_NN} summarizes these similarities and differences between both networks. 

\begin{table}[h]
\caption{Relationship of parameters between artificial and biological networks}
\label{table:comparison_CNN_NN}
\setlength\tabcolsep{2pt}
\resizebox{\columnwidth}{!}{
\begin{tabular}{|c|c|c|}
\hline

 & CNN & Simulation  \\ \hline

Number of neurons & \multicolumn{2}{c|}{276} \\ \hline
Number of layers & \multicolumn{2}{c|}{3}\\\hline
Neuronal topology & \multicolumn{2}{c|}{200 (Layer 1), 72 (Layer 2), 4 (Layer 3)} \\\hline
Input data & \multicolumn{2}{c|}{Maze} \\\hline
Types of neurons & Artificial & \begin{tabular}{@{}c@{}} Pyramidal neuron from \\ primary visual cortex \end{tabular} \\\hline
Connection weights & Filter weights & Synaptic weights \\\hline

\end{tabular}}
\end{table}

To represent the behavior of each neuron, we decided to use the Izhikevich neuronal model since it is computationally inexpensive, and it allows us to precisely model different types of neurons within different regions of the brain \cite{Izhikevich2003}. This model represents an abstraction of how cortical neurons behave in the brain. In particular, the following set of equations describes the Izhikevich model, whose parameters are indicated in Table~\ref{table:Izhikevich}. This model allows multiple configurations to mimic different regions of the brain. In our scenario, we assigned particular values to the previous parameters to implement a regular spiking signaling from the cerebral cortex, as indicated in \cite{Izhikevich2003}. Specifically, we aimed to model pyramidal neurons from the primary visual cortex of a mouse, which correspond to excitatory neurons typically present in the biological visual layers L2/3, L5, and L6 \cite{Bachatene2012}. For simplicity, during the analysis of the results of the simulation, we will refer to these layers in subsequent sections as first layer (L2/3), second layer (L5) and third layer (L6).

\begin{table}[h]
\caption{Parameters used in the Izhikevich model}
\label{table:Izhikevich}
\setlength\tabcolsep{2pt}
\resizebox{\columnwidth}{!}{
\begin{tabular}{|c|c|c|}
\hline

Parameter & Description & Values  \\ \hline

$v$ & Membrane potential of a neuron & [-65, 30] mV \\ \hline
$u$ & Membrane recovery variable providing negative feedback to $v$ & (-16, 2) mV/ms \\ \hline
$a$ & Time scale of $u$ & 0.02/ms \\ \hline
$b$ & Sensitivity of $u$ to the sub-threshold fluctuations of $v$ & 0.2/ms \\ \hline
$c$ & After-spike reset value of $v$ & -65mV \\ \hline
$d$ & After-spike reset value of $u$ & 8mV/ms \\ \hline
$I$ & Injected synaptic currents & \{10, 15\} mV/ms \\ \hline

\end{tabular}}
\end{table}

\noindent
\begin{equation} \label{eq:izhikevich1}
        v'=0.04v^2+5v+140+u+I
\end{equation}

\noindent
\begin{equation} \label{eq:izhikevich2}
        u'=a(bv-u) 
\end{equation}

\noindent
\begin{equation} \label{eq:izhikevich3}
        if v \geqslant 30mV, then \begin{cases}
          v \leftarrow c \\
          u \leftarrow u+d 
        \end{cases}      
\end{equation}
        
\begin{figure}[!h]
    \centering
    \begin{subfigure}[b]{\columnwidth}
        \centering
        \includegraphics[width=0.5\columnwidth]{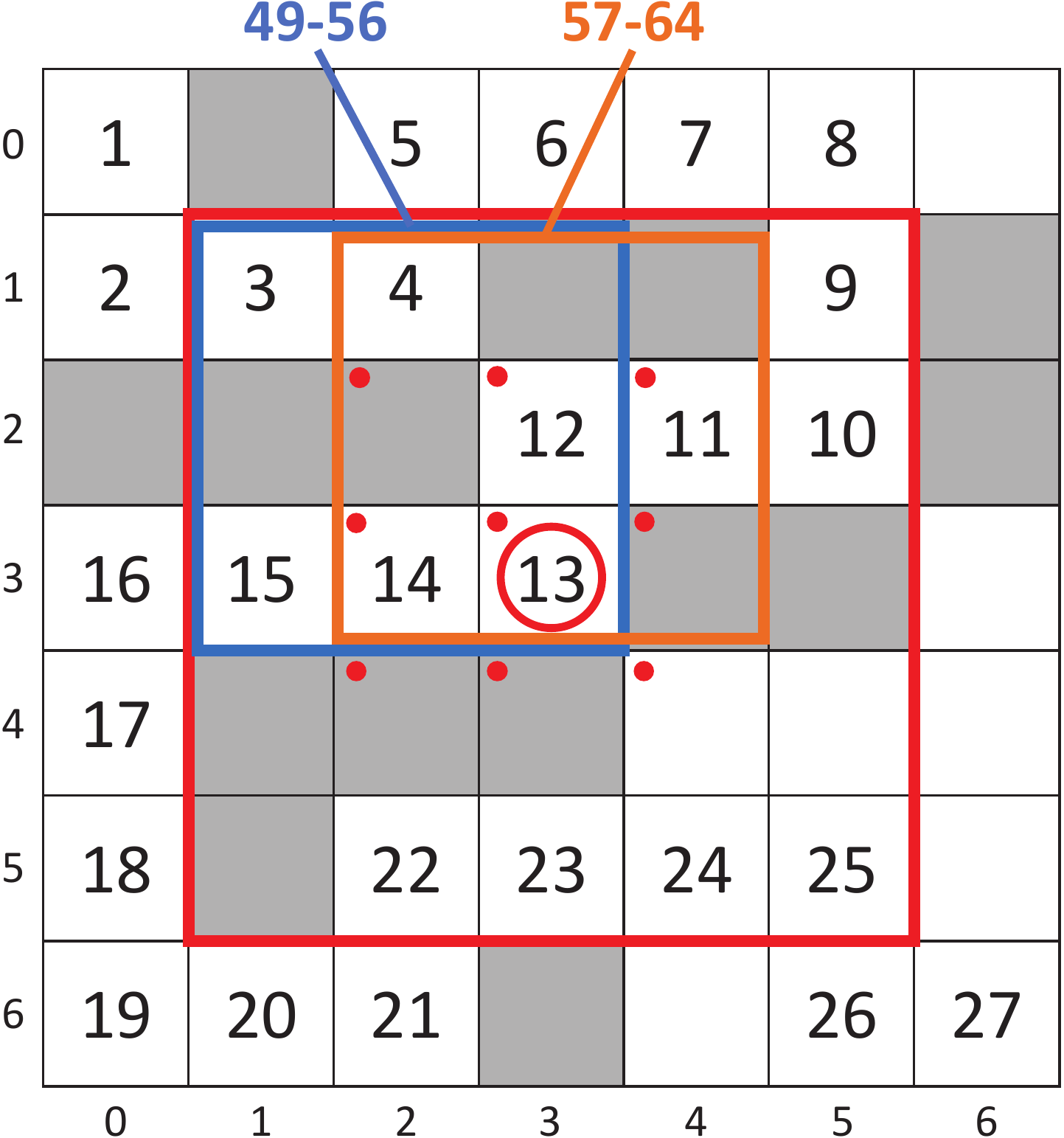}
        \caption{Relationship between visible positions from the current location of the mouse and their intervening neurons. In this example, the mouse is placed in position 13. Red dots indicate the visible positions from position 13.}
        \label{subfig:positions_neurons}
    \end{subfigure}
    %\hfill
    \begin{subfigure}[b]{\columnwidth}
        \includegraphics[width=\columnwidth]{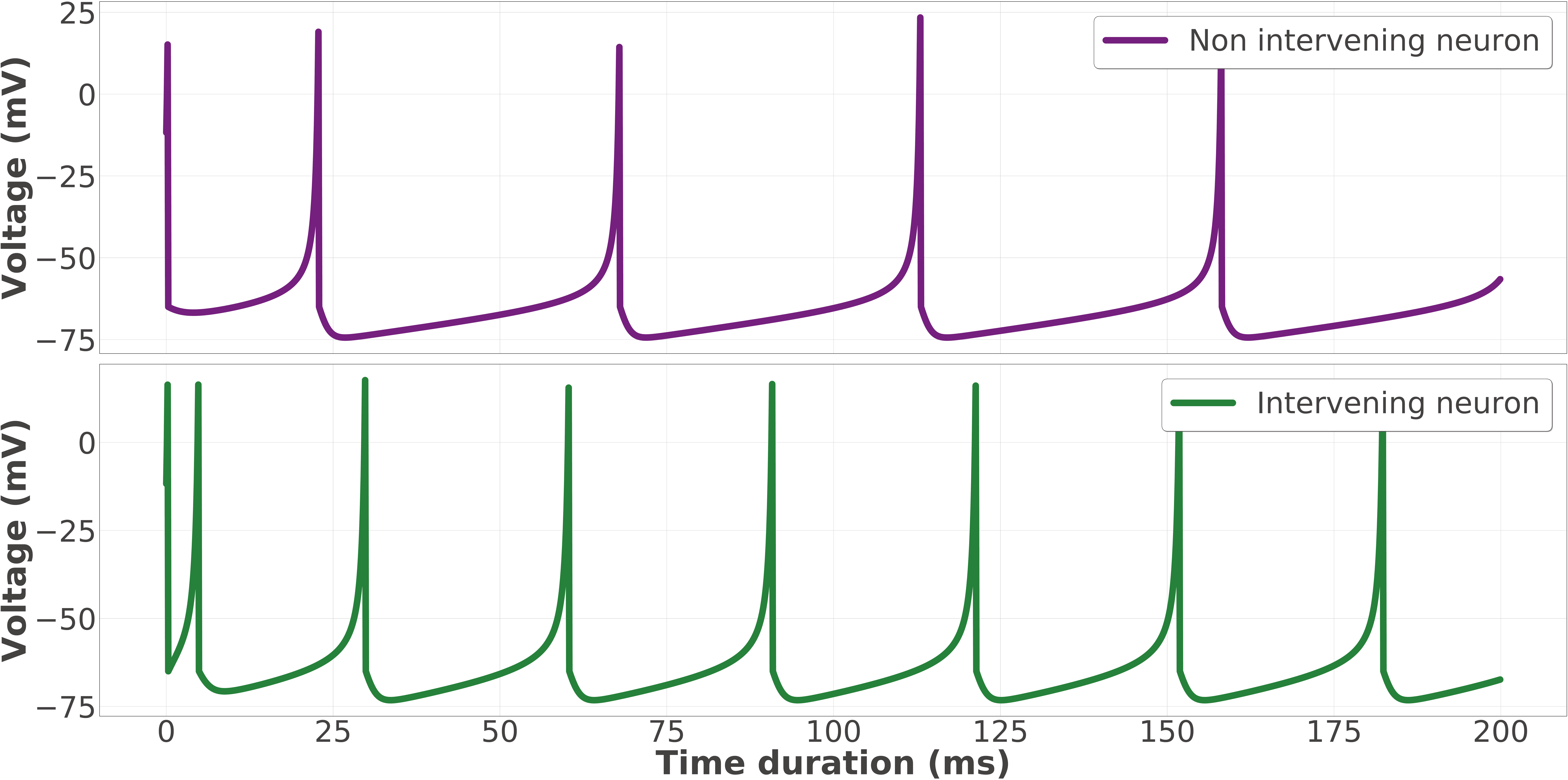}
        \caption{Impact of the I parameter on both intervening ($15mV$) and non intervening ($10mV$) neurons.}
        \label{subfig:differences_I}
    \end{subfigure}

    \caption{Relationship between positions of the maze and its implication in the modulation of neuronal signaling.}
    \label{fig:bots-characterization}
\end{figure}

To create our neuronal topology, we used the weights of the trained CNN as post-synaptic voltage values, normalized within the range between $5mV$ and $10mV$. We selected this range because these values constitute a conservative voltage raise within the range of values of $v$, indicated in Table~\ref{table:Izhikevich}. At the beginning of the simulation, we assigned the initial voltage of each neuron from a previously generated random list in the range $[-65mV, 0mV)$. This initial value for each neuron is constant between executions to allow their comparison. To define a more realistic use case, we represented in our simulation the movement of the mouse inside the maze (see \figurename~\ref{fig:mazeLayout}), staying one second in each position of the optimal path. To understand this, it is essential to introduce the concept of \textit{intervening neurons}, which defines the set of neurons managing all the visible positions of the mouse when it is placed in a particular position of the maze. \figurename~\ref{subfig:positions_neurons} illustrates the relationship between the position 13 of the optimal path and its intervening neurons, not considering its related visible positions for simplicity. For this position, we define nine 3$\times$3 squares within the surface delimited by the red square, where we represent only the first two squares to improve the legibility of the figure. Focusing on the first square, colored in blue, it comprises eight neurons indexes (49 to 56), obtained from the translation between 3-dimension coordinates previously commented in this section. The second one, highlighted in orange, associates eight different neurons. After applying all nine squares, we obtain the complete list of intervening neurons related to the position 13. This single process is repeated for every visible position from the position 13 (indicated in \figurename~\ref{subfig:positions_neurons} with red dots), obtaining the complete set of intervening neurons. This set of intervening neurons is presented in Table~\ref{table:intervening_neurons}, where each visible position from the position 13 is identified by its maze coordinate for simplicity. The last row of the table presents the complete set of intervening neurons for the position 13, obtained as the union of all individual sets of neurons. 

The movement of the mouse was implemented by providing external stimuli to the simulation via the $I$ parameter, where a value of $15mV$ was assigned to all intervening neurons from the current location of the mouse. For all non-intervening neurons in a specific instant, we assigned a value of $10mV$. These values align with the range defined in \cite{Izhikevich2003}. This information was extracted from the topology of the CNN, which contains the relationship between the neurons of the first layer and the positions of the maze. We took into consideration these aspects in the experimental analysis performed in Section~\ref{sec:results_analysis}. Based on that, we modeled with a higher value of $I$ those intervening neurons, transmitting a more potent visual stimulus to those neurons related to adjacent positions from the current location. Based on Equation \ref{eq:izhikevich1}, an increase in the $I$ parameter will produce a voltage rise in these intervening neurons, generating a raise in the amplitude of the electrical signal. This behavior was modeled taking into consideration the study performed in \cite{norcia:ssvep:2015}, which indicates that a known visual stimulus generates a voltage amplitude increase. \figurename~\ref{subfig:differences_I} graphically compares these differences between values of the $I$ parameter. It highlights that intervening neurons present a higher number of spikes during a particular temporal window, which is interpreted by the brain as the reconnaissance of accessible cells in the maze from the current position.

\begin{figure*}[h]
\begin{center}
\includegraphics[width=0.7\textwidth]{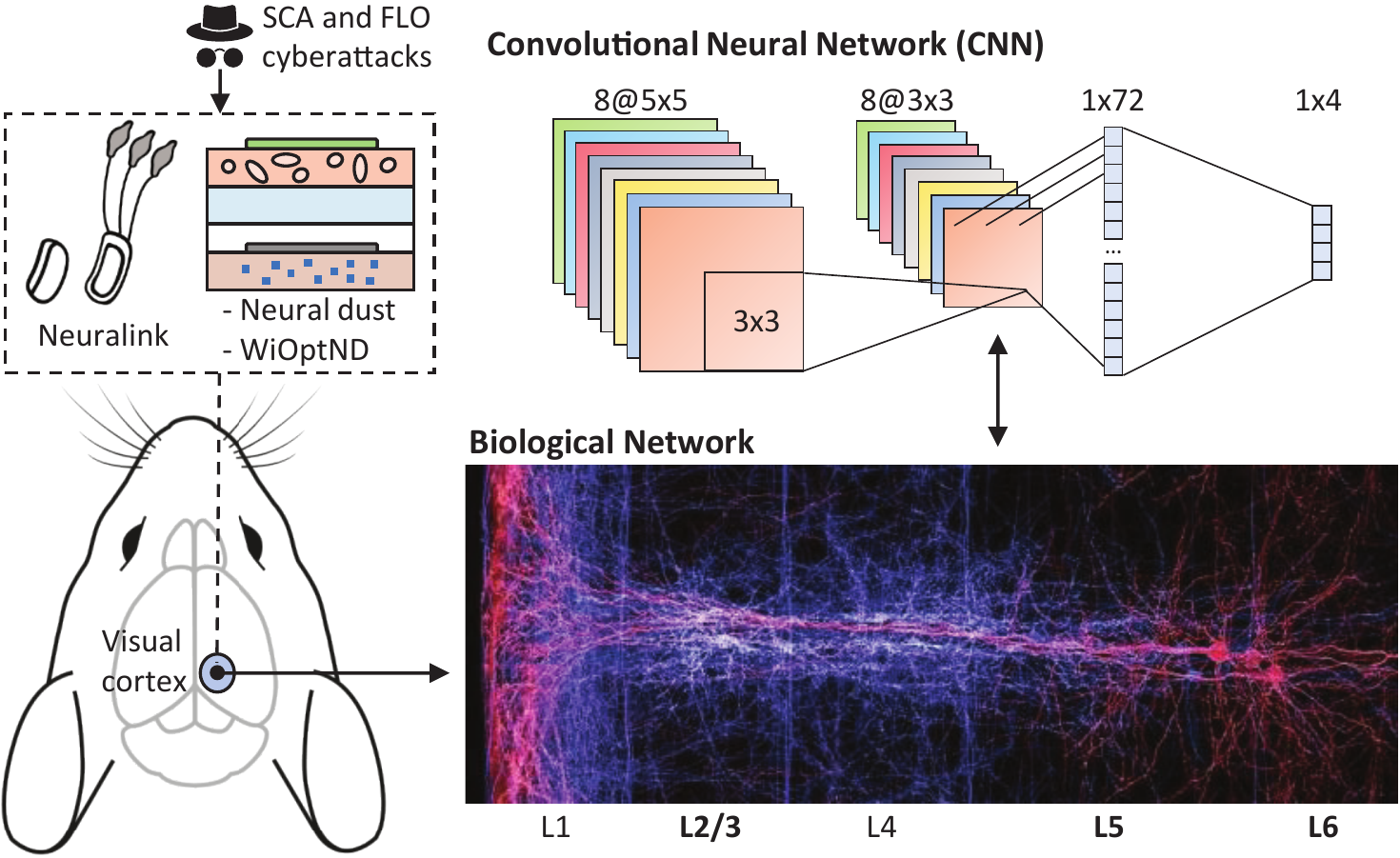}
\end{center}
\caption{Summary of our use case, indicating the translation between the topology of the CNN implemented and the biological network simulated.}
\label{fig:comparative_networks}
\end{figure*}

Finally, \figurename~\ref{fig:comparative_networks} introduces a graphical summary of the current use case. It depicts a mouse with a miniature brain implant solution in its primary visual cortex, such as Neuralink or Neural dust. To simulate its biological neuronal network, and based on a lack of realistic cortical topologies, a trained CNN provides the number of nodes and distribution in layers for the biological network. In particular, we modeled pyramidal neurons from visual layers L2/3, L5, and L6, using the Izhikevich model with a regular spiking signaling. Based on this scenario, an external attacker takes advantage of contemporary vulnerabilities in these implantable solutions to alter the behavior of the spontaneous activity of the biological neuronal network.

\begin{table}[h]
\caption{List of intervening neurons associated to the position 13 of the optimal path of the maze.}
\label{table:intervening_neurons}
\setlength\tabcolsep{2pt}
\centering
\begin{tabular}{|c|c|}
\hline

Coordinate & List of intervening neurons  \\ \hline

(2,2) & [1,24], [41,64], [81,104] \\ \hline
(2,3) & [9,32], [49, 72], [89, 112] \\ \hline
(2,4) & [17,40], [57,80], [97, 120] \\ \hline
(3,2) & [41,64], [81, 104], [121,144] \\ \hline
(3,3) & [49,72], [89,112], [129,152] \\ \hline
(3,4) & [57,80], [97,120], [137,160] \\ \hline
(4,2) & [81,104], [121,144], [161,184] \\ \hline
(4,3) & [89,112], [129,152], [169,192] \\ \hline
(4,4) & [97,120], [137,160], [177,200] \\ \hline \hline
Position 13 & [1,200] \\ \hline

\end{tabular}
\end{table}

%% file: tex/6results.tex
\section{Results analysis based on metrics}
\label{sec:results_analysis}

In this section, we evaluate the impact that FLO and SCA cyberattacks have on spontaneous neuronal activity of the neuronal topology presented in Section~\ref{sec:use_case}. To analyze the evolution of the cyberattacks impact while the mouse is moving across the maze, we consider the following three metrics:

\begin{itemize}
    \item Number of spikes: determine if a cyberattack either increases or reduces the quantity of spikes compared to the spontaneous neuronal signaling.
    \item Percentage of shifts, being a shift the delay of a spike in time (forward or backward) compared to the spontaneous behavior: study if a cyberattack generates significant delays in the normal activity of the neurons.
    \item Dispersion of spikes in both dimensions of time and number of spikes: analyze the spiking patterns under attack, aiming to detect if the cyberattack causes a modification on the distribution of the spikes.
\end{itemize}

For each layer of the topology, and combining all of them, we measured and analyzed the number of spikes and percentage of shifts. Finally, the dispersion of spikes is computed for each position of the optimal path and grouping all layers. Finally, we compared the impact generated by both cyberattacks. 

\begin{figure*}[h]
\begin{center}
\includegraphics[width=\textwidth]{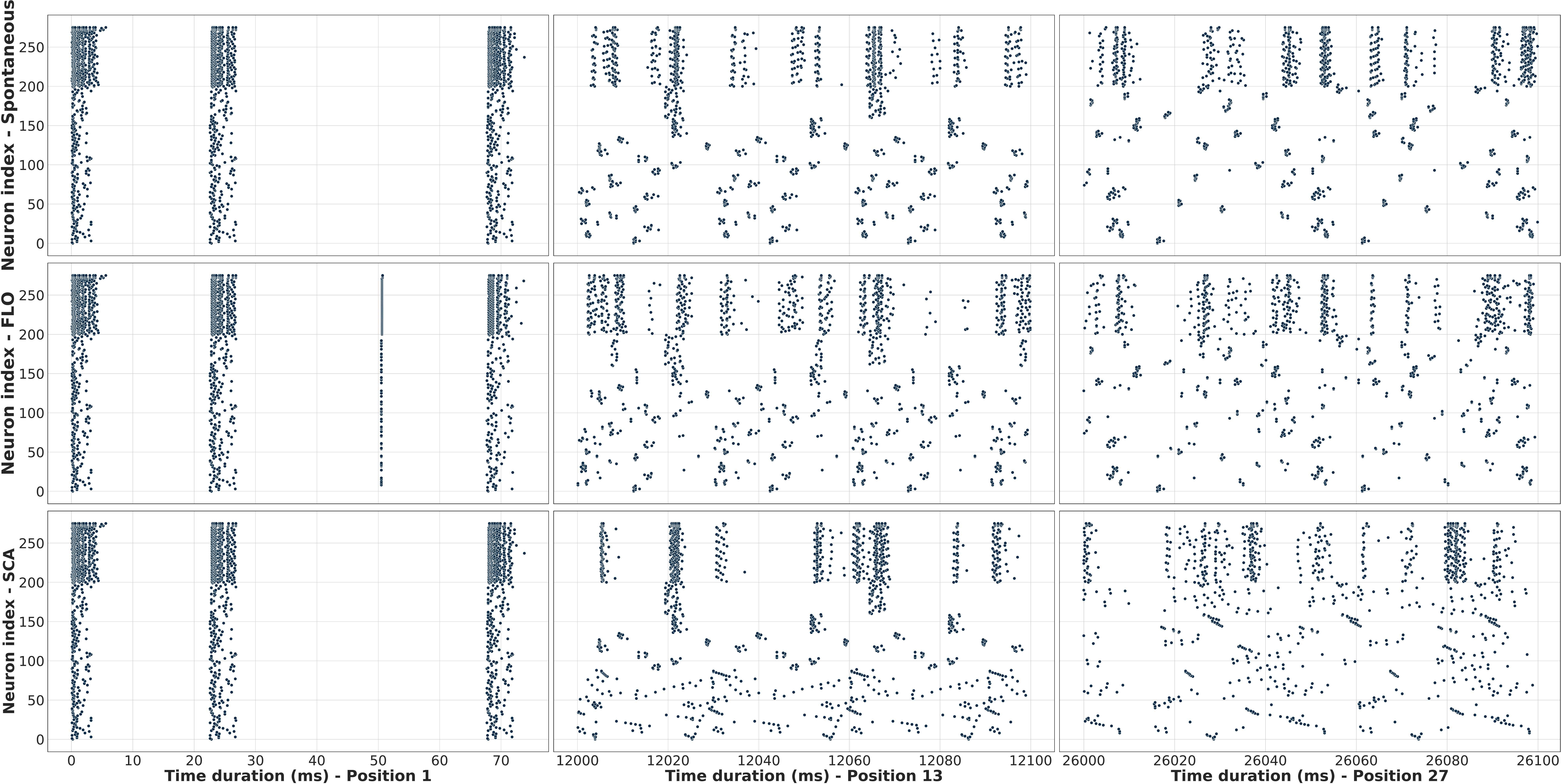}
\end{center}
\caption{Raster plots indicating the evolution of the spontaneous signaling and both FLO and SCA cyberattacks for three positions of the optimal path of the maze.}
\label{fig:raster_plots}
\end{figure*}

To better understand the impact of FLO and SCA cyberattacks, \figurename~\ref{fig:raster_plots} compares the evolution of neuronal spikes for the spontaneous activity, a FLO cyberattack and an SCA cyberattack. We selected three positions of the optimal path to analyze in detail the spiking evolution along with the simulation, presenting only the first 100ms of each position. It is essential to note that this simplification is only for this figure, and all the results subsequently presented consider the complete duration of each position. As can be seen, in the spontaneous signaling, there is a certain natural dispersion caused by the behavior of the neuronal model used, and the movement of the mouse (due to the the modification of the associated $I$ parameter). Specifically, each time the mouse changes from one position to another, the $I$ parameter changes according to the intervening neurons, where a higher value of $I$ is translated to a higher spike rate (see Algorithm \ref{alg:formalFLO}). Since the mouse periodically changes its position, it modifies the spiking rate of the neurons, generating a natural dispersion in the absence of attacks. Looking at the first position of both spontaneous and FLO, in the instant 50ms, there is a clear difference between them, since we executed the attack in that exact instant. The set of attacked neurons generates spikes before it was intended due to the voltage rise produced by the attack. Consequently, we can see that the dispersion over the following positions (13 and 27) augments, altering the natural pattern of the neurons. Regarding the SCA cyberattack, it also starts in the instant 50ms but, its impact it is not yet present in the first 100ms of the initial position. If we check the subsequent positions, the attack gradually propagates, generating characteristic ascending patterns. Subsequent subsections analyze, in a more detailed way, the information contained in \figurename~\ref{fig:raster_plots}, extending the analysis to all the positions of the optimal path and using the previous three metrics. 

\subsection{Neuronal Flooding}

In this subsection, we aim to simultaneously attack multiple neurons and analyze its impact using the metrics previously indicated at the beginning of the section. The implementation of this cyberattack is based on the general description indicated in Algorithm \ref{alg:formalFLO}. We decided to perform only the attacks over the first layer of the topology, from where each target neuron is randomly selected, to evaluate the propagation to deeper layers. Furthermore, we tested a combination of two additional parameters. The first one represents the number of simultaneously attacked neurons, 
%$|\mathbb{AN}|
$k \in \{5,15,...,95,105\}$. $\mathbb{AN}$ will contain $k$ neurons randomly selected from $\mathbb{NE}$, the set of neurons in the first layer. 
It is worthy to note that we reached to attack simultaneously more than half of the neurons of the first layer, which represents a fairly aggressive portion of the neurons. The second parameter of the attack, $\mathbb{VI}=\{20,40,60\}$, indicates the different voltage increases in $mV$ used to stimulate the neurons in $\mathbb{AN}$. Its maximum level, $60mV$, approximately represents two-thirds of the voltage range defined by the Izhikevich model. We have executed each combination of parameters 10 times, denoted as $exec = 10$, to ensure that the random selection of neurons performed is representative. The value of $\textbf{t}^\text{sim}$ is 27s (one second per position of the optimal path), and $\textbf{t}^\text{attk}$, is 50ms. Table~\ref{table:parametersFLO} summarizes the previously indicated parameters. 

\begin{table}[h]
\caption{Configuration of the implemented FLO cyberattack}
\label{table:parametersFLO}
\setlength\tabcolsep{2pt}
\resizebox{\columnwidth}{!}{
\begin{tabular}{|c|c|c|}
\hline

Parameter & Description & Values  \\ \hline
$\textbf{t}^\text{sim}$ & Duration of the simulation & 27s \\ \hline
$\textbf{t}^\text{attk}$ & Instant of the attack & 50ms \\ \hline
$\mathbb{NE}$ & Set containing all neurons of the first layer & $\{1,2,...,199,200\}$ \\ \hline
$|\mathbb{AN}|$ & Number of simultaneously attacked neurons & $\{5,15,...,95,105\}$ \\ \hline
$\mathbb{VI}$ & Set containing the voltage used to attack random neurons & $\{20,40,60\}mV$ \\ \hline
$exec$ & Number of executions per combination of parameters & $10$ \\ \hline

\end{tabular}}
\end{table}

\subsubsection{Number of spikes metric}
\label{subsubsec:FLO_number_spikes}

\begin{figure}[h]
\begin{center}
\includegraphics[width=\columnwidth]{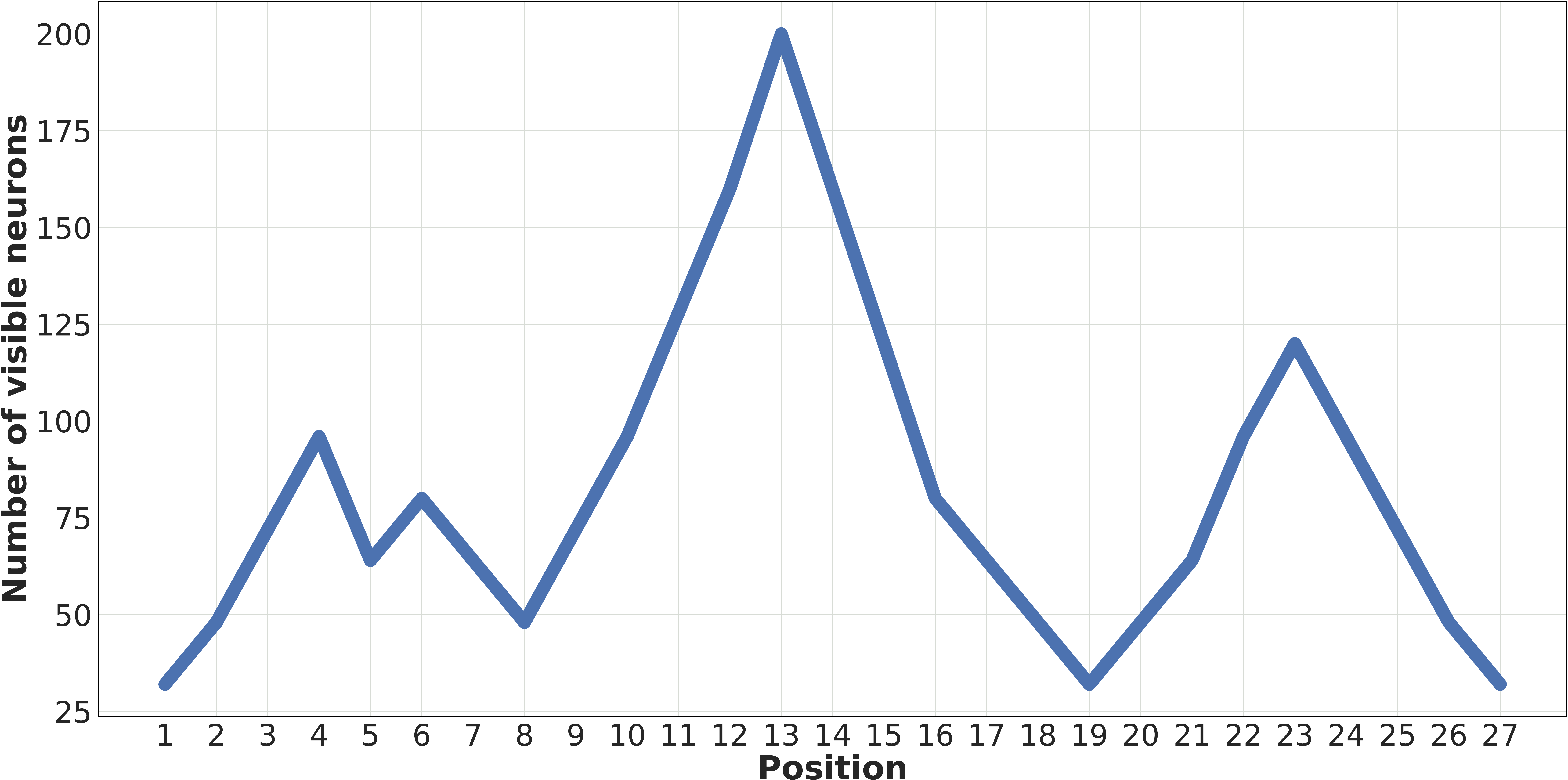}
\end{center}
\caption{Number of intervening neurons related to visible positions from each position of the optimal path.}
\label{fig:visible_neurons}
\end{figure}

To better understand the analysis of this metric, it is necessary to introduce \figurename~\ref{fig:visible_neurons}, which shows, for each position of the optimal path of the maze, the number of intervening neurons involved in the decision-making process of the mouse. Since these intervening neurons are dependent on the number of visible positions from a particular location of the maze, the number of intervening neurons is higher in central cells of the maze compared to those placed near the borders. Moreover, intervening neurons are dependent on the topology used and the convolution process of the CNN, as depicted in \figurename~\ref{fig:CNN_topology}.

\begin{figure}[h]
\begin{center}
\includegraphics[width=\columnwidth]{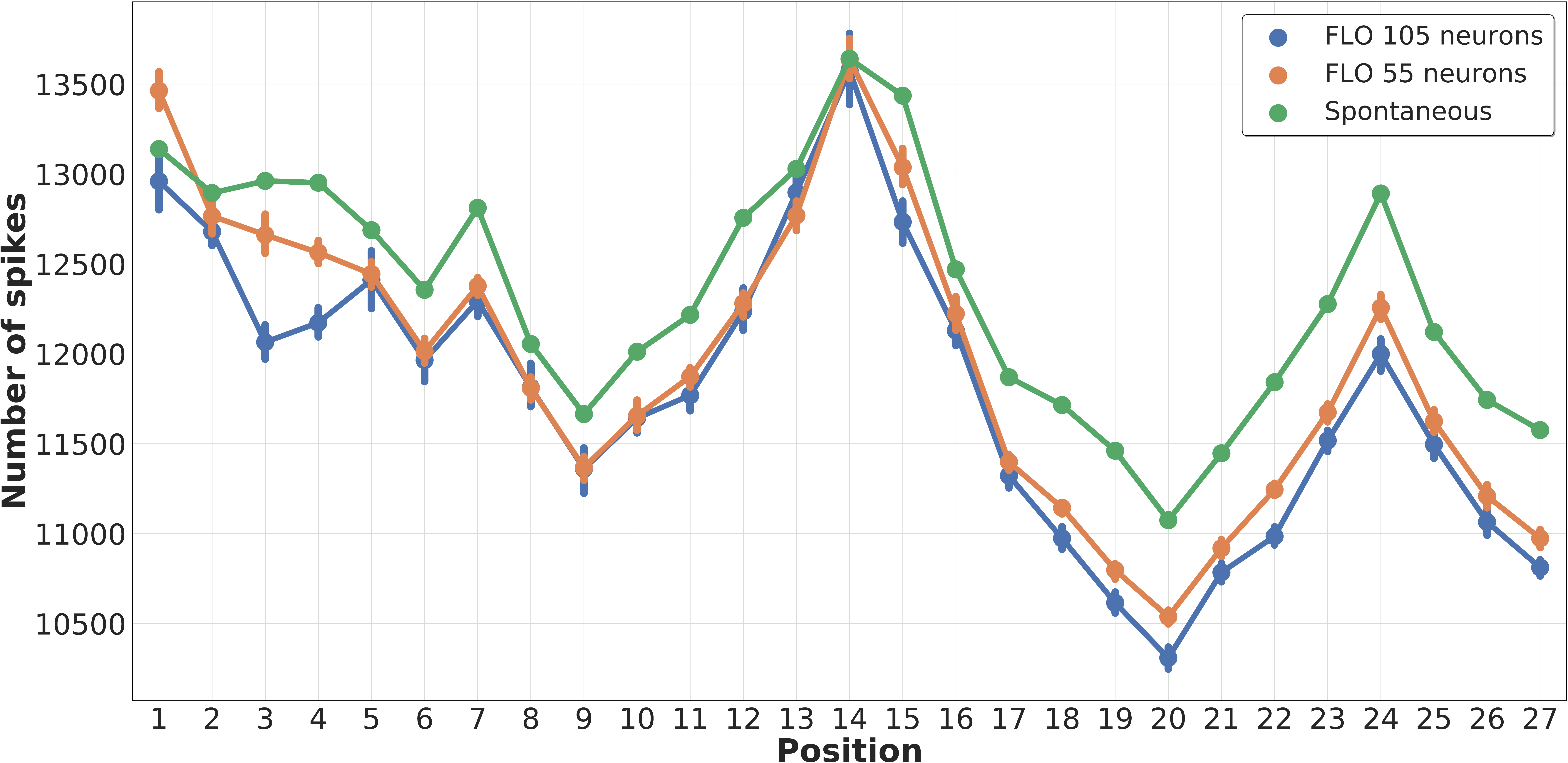}
\end{center}
\caption{Total number of spikes for all neurons of the topology per position of the optimal path, attacking different number of neurons (105 and 55 simultaneous neurons).}
\label{fig:FLO_total_spikes}
\end{figure}

\figurename~\ref{fig:FLO_total_spikes} compares, for the spontaneous signaling and two different configurations of FLO, the total number of spikes per position of the optimal path. In particular, the graph plots two different amounts of neurons in $\mathbb{AN}$ (55 and 105 neurons) for all $exec$ simulations. In this figure, we fixed $v_i$ to a value of $40mV$ to improve its visualization. As can be seen, both figures share a common tendency, indicating that the higher the number of intervening neurons from a position, the higher the number of spikes. This is a consequence of how the mouse moves across the maze and how neurons and positions are related based on our particular topology. Comparing both figures, \figurename~\ref{fig:visible_neurons} reaches its highest peaks one position before, since this change of intervening neurons needs to be propagated in time, affecting the number of spikes of its following position.

In \figurename~\ref{fig:FLO_total_spikes}, we can see that, in general, FLO cyberattacks reduce the number of spikes compared to the spontaneous activity, increasing this reduction when the mouse progresses in the maze. Furthermore, increasing the impact of the attack, in terms of the number of attacked neurons, reduces the number of spikes. These aspects are aligned with the results later presented in Section~\ref{subsubsec:FLO_dispersion}, where this reduction is caused by an increase of the dispersion in the attacked neurons. However, it is worth noticing the high number of spikes produced in the first position. The Izhikevich neuronal model for regular spiking generates a quick burst of spikes in a short time, and, after that, it stabilizes its spike rate, explaining this behavior. When we apply a FLO cyberattack, the attacked neurons anticipate their spikes, producing either a raise of spikes if the number of attacked neurons is not so elevated (low dispersion in time), or a reduction of spikes if most of the neurons are attacked (high dispersion). Moreover, the evolution of the simulation after the attack does not tend to come back to the spontaneous signaling, in terms of the number of spikes. In fact, these distances augment over time, reaching a difference of around 700 spikes in position 27, with some variability between both FLO configurations. Based on that, these results indicate that the effect of attacking neurons in a particular instant propagates until the end of the simulation.

\begin{figure}[h]
\begin{center}
\includegraphics[width=\columnwidth]{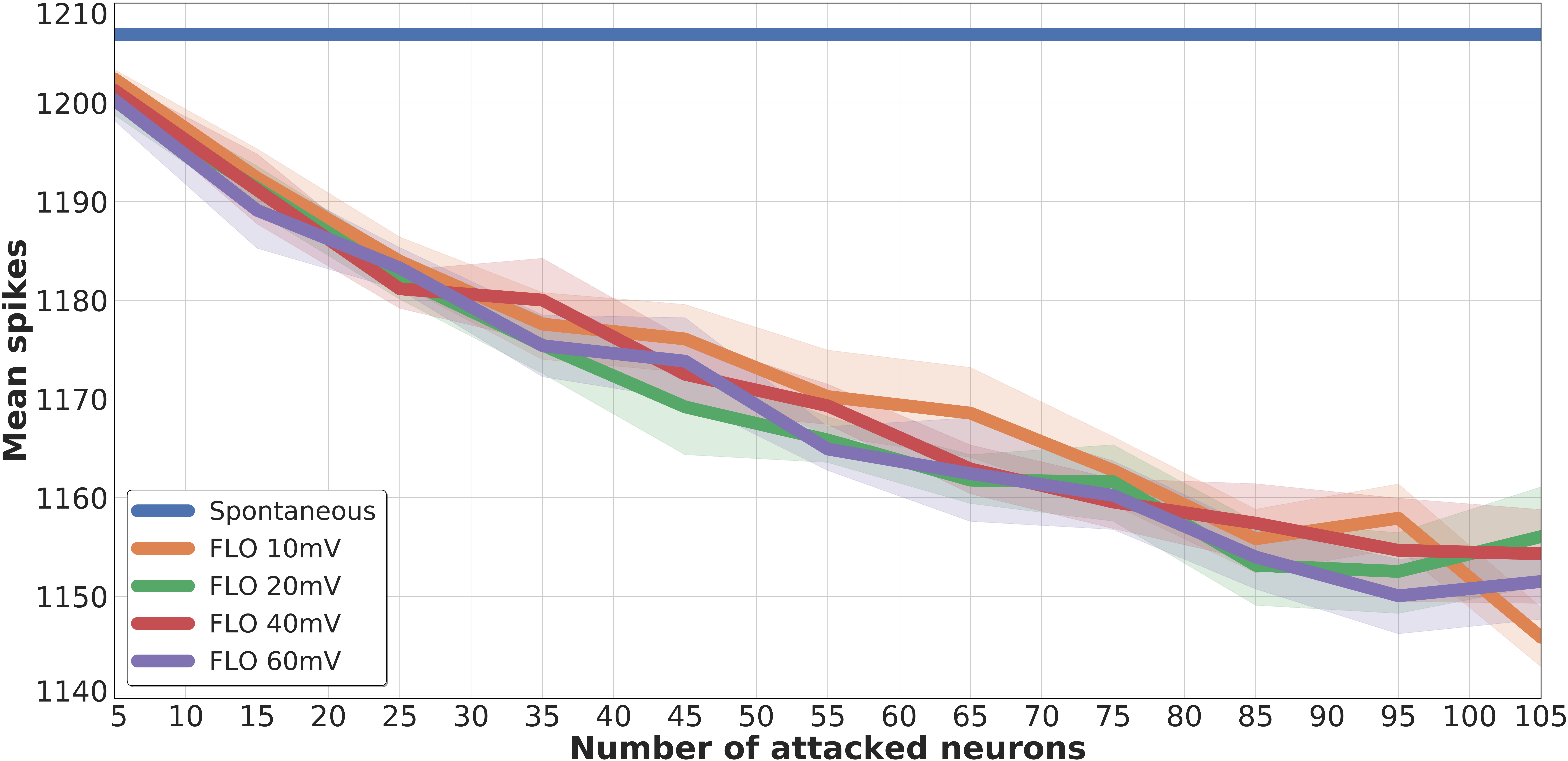}
\end{center}
\caption{Evolution of the mean of spikes with different number of attacked neurons and voltage increases, aggregating all positions of the optimal path.}
\label{fig:FLO_global_spikes}
\end{figure}

After this analysis, we considered relevant to evaluate how the mean of spikes evolved through the three layers of the topology with different configurations of the FLO cyberattack. In particular, we tested different amounts of attacked neurons and voltage increase, with $exec$ different executions for each combination of the previous parameters. Using $exec$ executions introduces variability in terms of the randomly selected neurons for each execution. We present these results in \figurename~\ref{fig:FLO_global_spikes}, which represents an aggregation of the number of spikes produced during the optimal path of the maze. It indicates that increasing the number of attacked neurons derives in a higher reduction in the number of spikes, while the application of different voltages does not produce a high impact. The dimmed colors surrounding the main lines of the figure indicate the fluctuations between the $exec$ simulations. As can be seen, the difference in the mean of spikes compared to the spontaneous signaling grows when the number of attacked neurons raises, having a difference of around 60 spikes for 110 attacked neurons (half of the first layer). These results align with those presented in \figurename~\ref{fig:FLO_total_spikes} for the positions of the optimal path, where both figures present a clear descending trend when the number of attacked neurons augments. Finally, the use of different increases of voltage during the experiments did not generate a considerable impact on the number of spikes.

\begin{figure}[h]
\begin{center}
\includegraphics[width=\columnwidth]{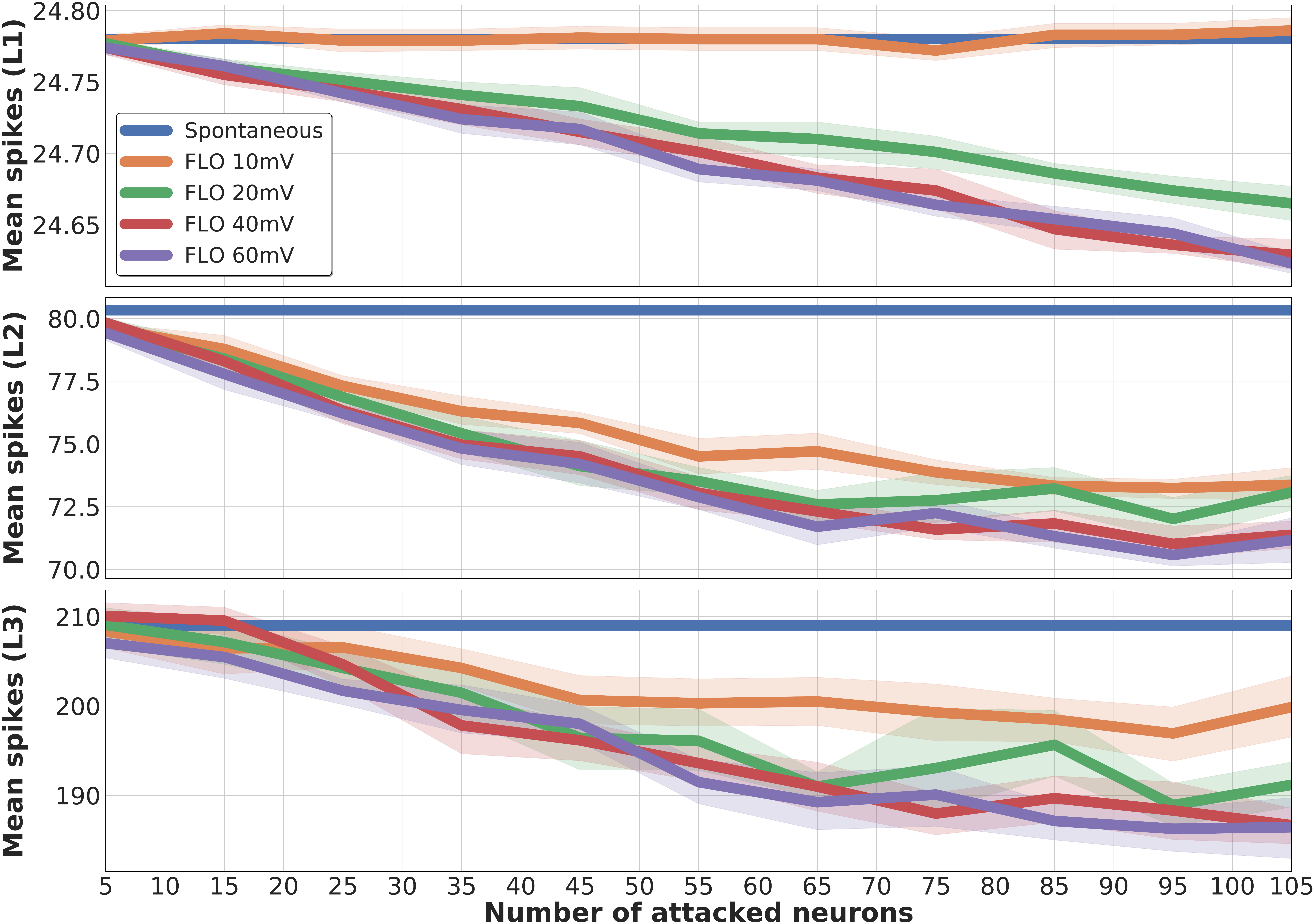}
\end{center}
\caption{Mean of spikes for each layer of the topology, focusing on the last position of the optimal path.}
\label{fig:FLO_layers_spikes}
\end{figure}

To expand the focus on this analysis and to determine whether this descending trend is exclusive to only certain layers, \figurename~\ref{fig:FLO_layers_spikes} analyzes the same parameters but differentiating between the three layers of the topology and focusing only on the last position of the optimal path of the maze. We can see that the variation of the mean of spikes is more significant in deeper layers (2nd and 3rd). This variation is due to the distribution of our topology and the normal behavior of the brain, where initial layers propagate their behavior to subsequent layers, magnifying their activity via synapses. The y-axis range considerably differs between layers, being the difference with the spontaneous signaling of less than one spike in the first layer. The second layer offers a broader range of around 8 spikes in the most damaging situation, whereas the third layer has an approximate separation of between 10 to 25 spikes. 

In summary, the previous figures indicate that, under attack, the mean of spikes decreases compared to the spontaneous behavior. In particular, we highlight that increasing the number of attacked neurons derives in a higher impact in the mean of spikes. Nevertheless, there are no significant differences in the variation of the voltage used to attack the neurons. Finally, the number of intervening neurons from the visible positions of the optimal path of the maze strongly influences the mean of spikes.

\subsubsection{Percentage of shifts metric}
\label{subsubsec:FLO_shifts}

\begin{figure}[h]
\begin{center}
\includegraphics[width=\columnwidth]{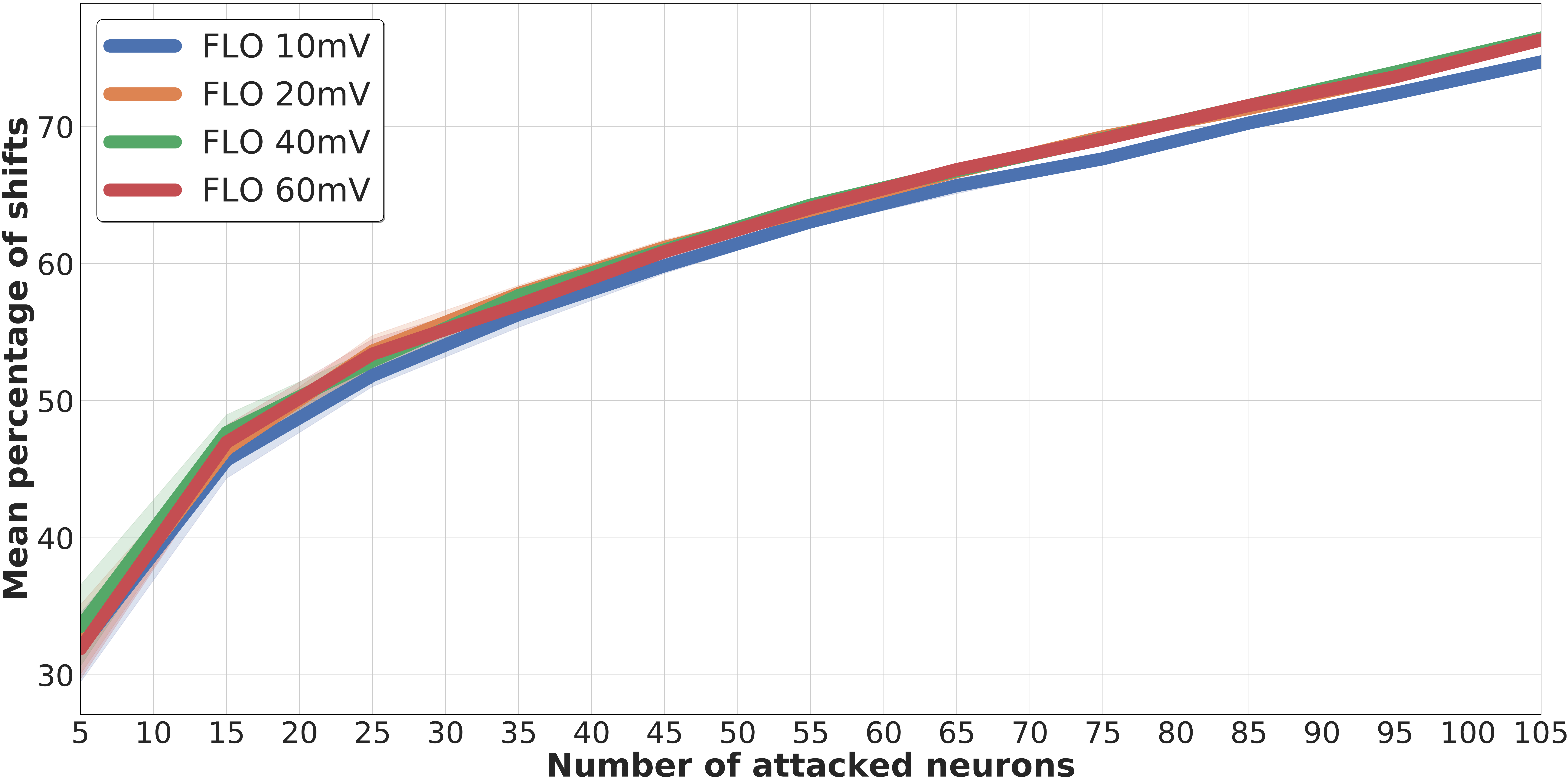}
\end{center}
\caption{Shift percentage mean for an aggregation of all topological layers and positions of the optimal path.}
\label{fig:FLO_global_shifts}
\end{figure}

For this metric, we first evaluated the percentage of delayed shifts for an aggregation of all three layers. After that, we analyzed the same but combining all the positions of the optimal path of the maze. In this test, we included a different number of attacked neurons and voltage raises. \figurename~\ref{fig:FLO_global_shifts} describes this situation, where attacking a higher number of neurons produces a higher percentage of shifts. This ascending trend is aligned with the dispersion metric, since an enlargement in the parameters of the attack produces a growth of shifts. As a consequence, it generates a higher dispersion in time and number of spikes.

\begin{figure}[h]
\begin{center}
\includegraphics[width=\columnwidth]{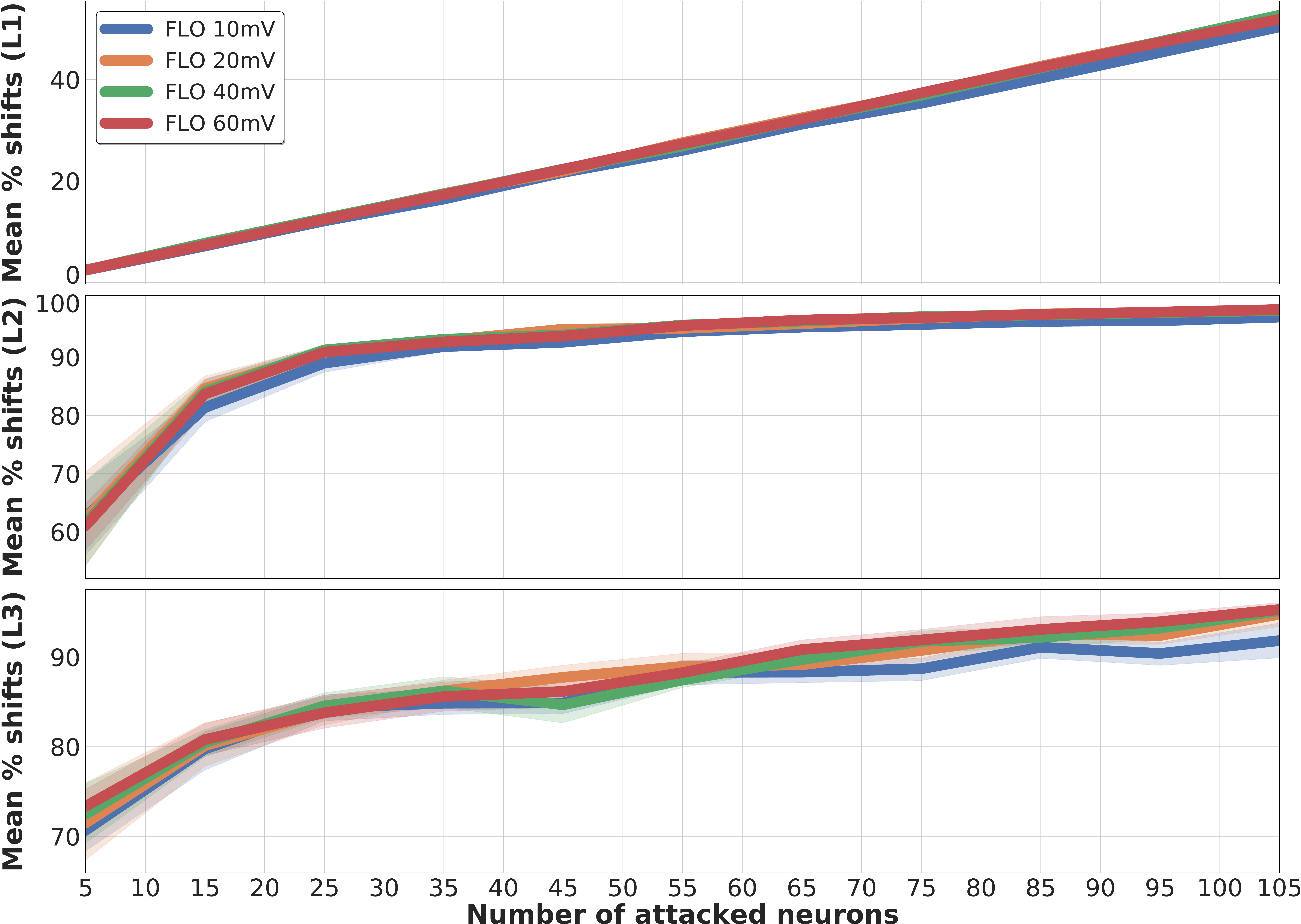}
\end{center}
\caption{Shift percentage mean for each layer of the topology, for the last position of the optimal path.}
\label{fig:FLO_layers_shifts}
\end{figure}

If we focus on each layer of the topology, \figurename~\ref{fig:FLO_layers_shifts} represents a FLO cyberattack for the last position of the optimal path, where each color line indicates a voltage raise. Focusing on the first layer, we can see a linear growth when we augment the number of attacked neurons since only those neurons shift in the layer. Moving to subsequent layers, we can observe that the growth tendency is more prominent in the second layer. This indicates that, when we advance to the third layer, the effect of the attack gets slightly attenuated. 

In conclusion, this metric indicates that attacking more neurons derives in a higher percentage of shifts. Additionally, and similarly to the metric studying the number of spikes, voltage increases have not a high impact on our scenario.

\subsubsection{Dispersion metric}
\label{subsubsec:FLO_dispersion}

We first focus on the spike dispersion over time caused by the different number of attacked neurons for each position of the optimal path. This means that, for each position of the maze, we obtain the number of time instants with recorded spikes, independently of the number of spikes. If we take into account that each position of the maze corresponds to one second and that the sampling rate of Brian2, by default, is $0.1ms$, we have a total number of $10\,000$ instants per position. If a position presents a higher dispersion value than other positions, it indicates that there are more instants with spikes in the former one. We focus on a voltage raise value of $40mV$, since previous analysis indicated that this parameter has a low impact on our scenario. 

\begin{figure}[h]
\begin{center}
\includegraphics[width=\columnwidth]{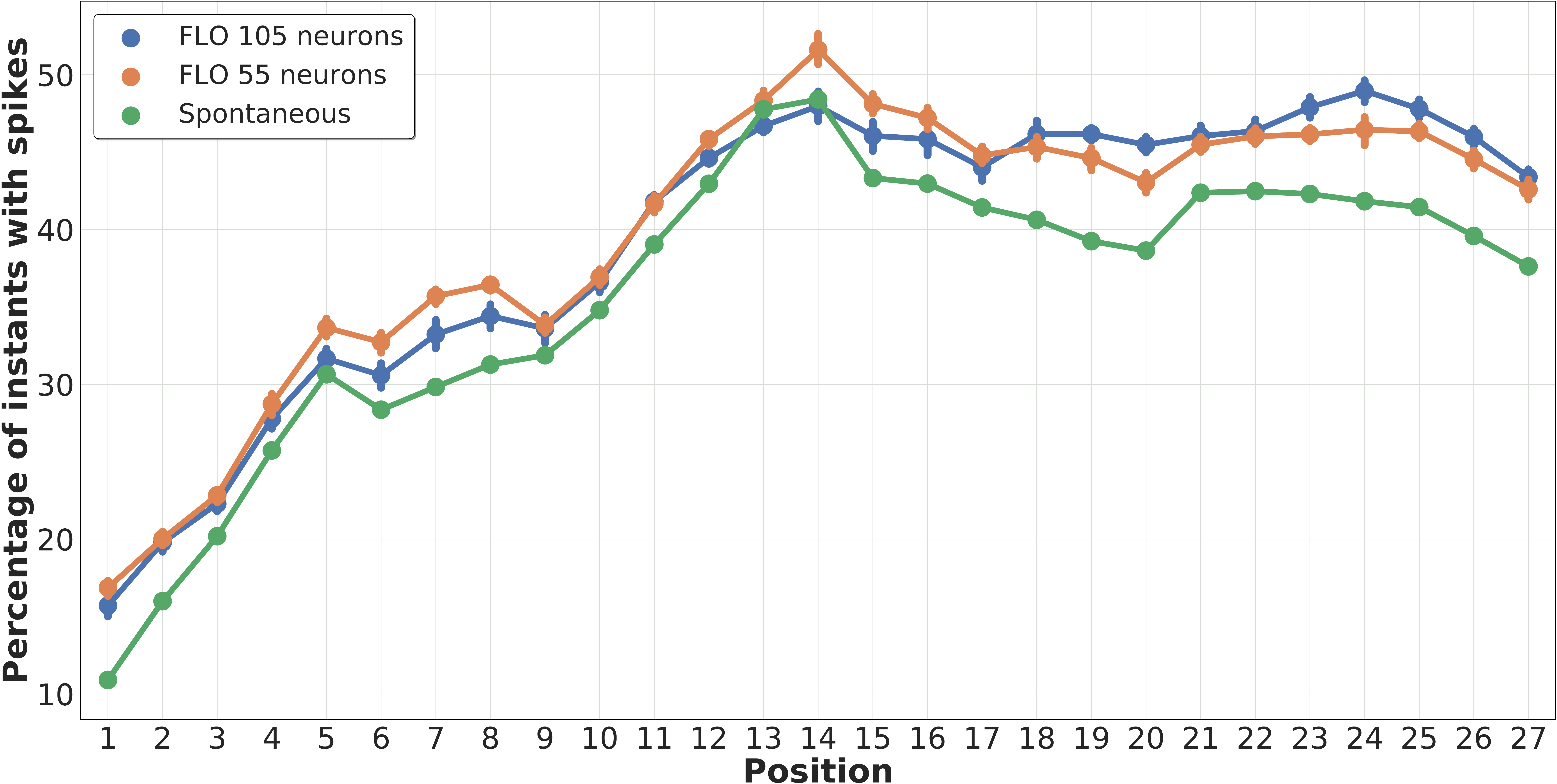}
\end{center}
\caption{Spike dispersion over time for each position of the optimal path.}
\label{fig:FLO_horizontal_dispersion}
\end{figure}

In \figurename~\ref{fig:FLO_horizontal_dispersion} we can observe that the spontaneous signaling presents some similarities with the trend existing in \figurename~\ref{fig:FLO_total_spikes} and, specifically, in those positions with the most significant peaks. If a position presents a raise in the number of spikes, the probability of having spikes in \figurename~\ref{fig:FLO_horizontal_dispersion} for a longer period of time also increases. However, the natural dispersion of the simulation attenuates these peaks, where the $I$ parameter changes according to the visible positions of the maze. Considering both FLO configurations, we can appreciate an enlargement in the temporal dispersion compared to the spontaneous behavior. FLO cyberattacks anticipate the spikes of the attacked neurons in a given moment, generating a higher dispersion as the simulation progresses. Specifically, the difference with the spontaneous signaling augments over time, induced by the natural variability of the mouse's movements. Although the attack with 55 neurons presents a higher impact until position 17, from that position until the end, the attack with 105 neurons has a higher impact from this metric. A higher impact over the temporal dispersion when we attack more neurons simultaneously aligns with the results presented in \figurename~\ref{fig:FLO_layers_spikes} for the number of spikes. These results are also related to those presented in Section \ref{subsubsec:FLO_shifts} for the percentage of shifts, where an intensification in these shifts derives in a dispersion growth.

\begin{figure*}[h]
\begin{center}
\includegraphics[width=\textwidth]{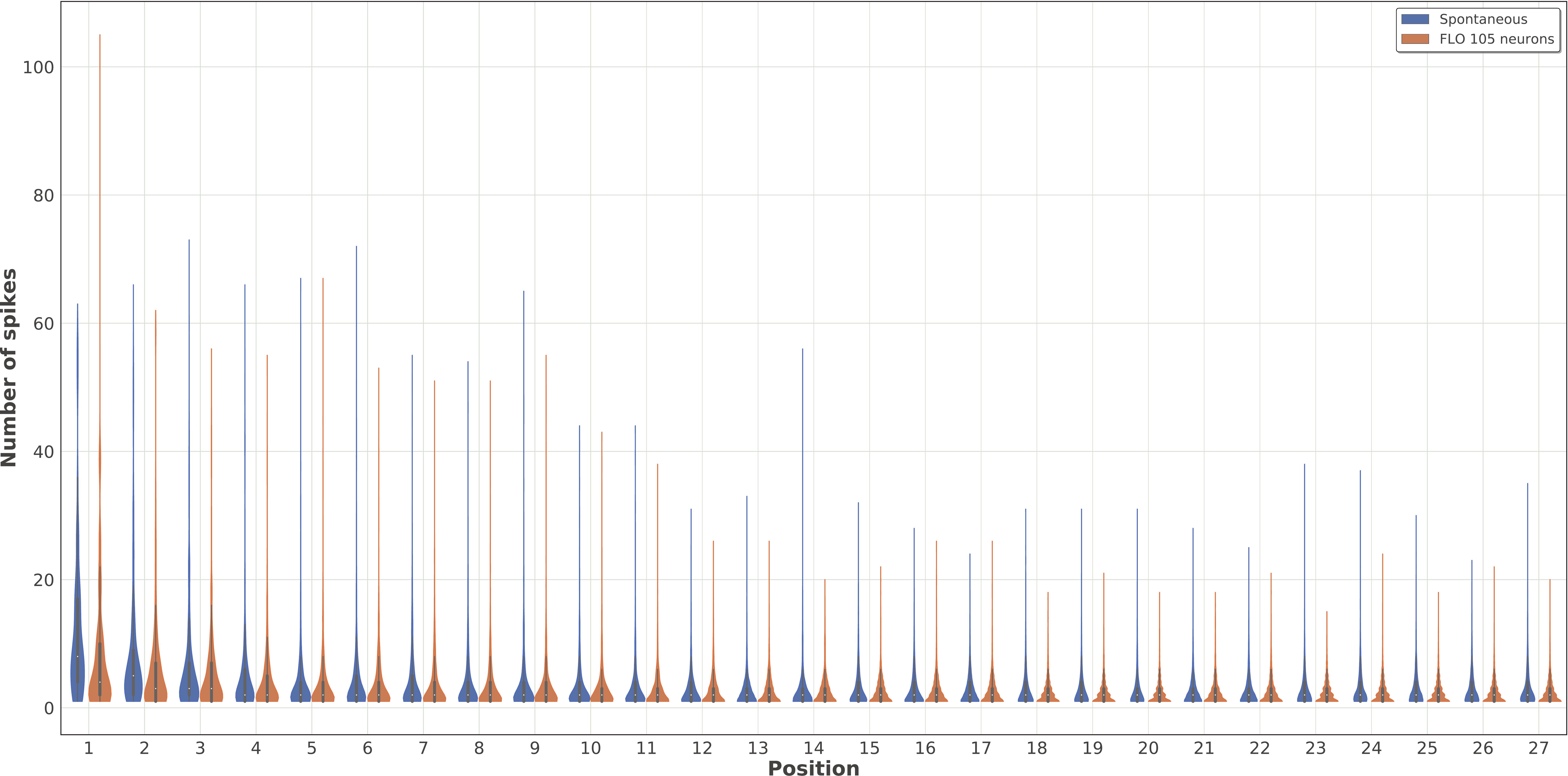}
\end{center}
\caption{Spike dispersion over the number of spikes for each position of the optimal path.}
\label{fig:FLO_vertical_dispersion}
\end{figure*}

We can also consider this dispersion from the perspective of the number of spikes. For each position of the optimal path, we evaluate the distribution of the number of spikes, setting the voltage increase to a value of $40mV$ and the number of simultaneous attacked neurons to 105. \figurename~\ref{fig:FLO_vertical_dispersion} illustrates this distribution, where each position contains a violin plot for both the spontaneous and under attack behaviors. It is essential to highlight that this figure represents only one of the $exec$ simulations performed for the complete set of experiments to ease the visualization. We can appreciate that the attack in position one reaches a peak of 110 spikes due to the increase of spikes induced by the attack performed at that particular moment. Focusing on the distribution indicated by each violin, the variance progressively reduces when the mouse progresses in the maze, concentrating the distribution of number of spikes around one. That means that in the last positions there are more instants where only one spike occurs, indicating that the attack increases the spike dispersion as the simulation progresses.

This situation aligns with the results presented in \figurename~\ref{fig:FLO_total_spikes}, where a higher number of spikes influence this upper threshold. Nevertheless, it is worth considering the exception in position 13, where this threshold is considerably reduced. To understand this situation, we also have to consider \figurename~\ref{fig:FLO_horizontal_dispersion}, which indicates that this position presents the highest percentage of dispersion, with more than 50\% of spikes shifted. This position indicates the relationship between these two dispersion approaches, where a high temporal dispersion generates a reduction in the dispersion focused on the number of spikes. 

In conclusion, FLO cyberattacks generate a large impact on the spontaneous neuronal activity. In particular, the previous figures highlight how the mouse's natural movement induces particular natural dispersion, both in time and number of spikes. Performing FLO cyberattacks also produces an enlargement in the temporal dispersion, where the neuronal activity is more scattered. This can also be analyzed from the dispersion focused on the number of spikes since this reduction on the aggregation causes the spikes to tend to a low number. It means that there are more instants with a fewer number of spikes compared to the spontaneous behavior.

% Comparative between all 3 metrics
The previous analysis, based on the number of spikes, percentage of shifts, and dispersion, highlights the impact that FLO cyberattacks can generate over the spontaneous neuronal activity. We subsequently analyze these metrics together since they are strongly dependent between them. In particular, the application of a FLO cyberattack generates a decrease in the number of spikes, where these differences are more prominent in deeper layers of the topology. These results can be explained based on the dispersion induced by the attack, where a growth on the dispersion reduces the probability of multiple action potentials in the first layer. Consequently, the post-synaptic voltage raises arrive at subsequent layers in a more dispersed way, delaying the spikes. The metric focused on the percentage of shifts over the spontaneous signaling is closely related to the dispersion metric. An increase in the percentage of shifts entails a modification in the natural periodicity of the spikes. This change is directly translated to a higher dispersion rate, both in time and number of spikes. Finally, it is essential to note that this behavior and results are dependent on our particular topology. Nevertheless, they can serve as an example of how performing a FLO cyberattack can affect neuronal activity in a particular scenario.

\subsection{Neuronal Scanning}

This section details the implementation of an SCA cyberattack on our topology, based on the general description of the attack represented by Algorithm~\ref{alg:formalSCA}. For this particular implementation, we have sequentially attacked the $200$ neurons that compose the first layer of the topology. We denote as $\mathbb{VI}=\{5,10,...,60,65\}$ the set of voltage raises, in $mV$, applied separately in each SCA cyberattack. As previously indicated for the FLO cyberattack, the duration of the simulation, $\textbf{t}^{sim}$, is 27s, staying the mouse one second in each position of the optimal path of the maze. Additionally, the attack initiates in the instant 50ms, represented by $\textbf{t}^{attk}$. To model the periodicity of attacking the neurons, $\Delta t$ indicates the temporal separation between two attacks over two consecutive neurons, being 134ms in our particular implementation. Each combination of parameters is executed only once ($exec = 1$) since there is no variability in the selection of neurons, as it is the case of a FLO cyberattack. Finally, Table~\ref{table:parametersSCA} indicates a summary of the parameters used in the implementation of SCA cyberattacks.

\begin{table}[h]
\caption{Configuration of the implemented SCA cyberattack}
\label{table:parametersSCA}
\setlength\tabcolsep{2pt}
\resizebox{\columnwidth}{!}{
\begin{tabular}{|c|c|c|}
\hline

Parameter & Description & Values  \\ \hline
$\textbf{t}^\text{sim}$ & Duration of the simulation & 27s \\ \hline
$\textbf{t}^\text{attk}$ & Start of the attack & 50ms \\ \hline
$\textbf{t}^\text{step}$ & Temporal duration between attacking two neurons & 134ms \\ \hline
$\mathbb{NE}$ & Set containing all neurons of the first layer & $\{1,2,...,199,200\}$ \\ \hline
$\mathbb{VI}$ & Set containing the voltage used to attack the neurons & $\{5,10,...,60,65\}mV$ \\ \hline

\end{tabular}}
\end{table}

\subsubsection{Number of spikes metric}
\label{subsubsec:SCA_number_spikes}

\begin{figure}[h]
\begin{center}
\includegraphics[width=\columnwidth]{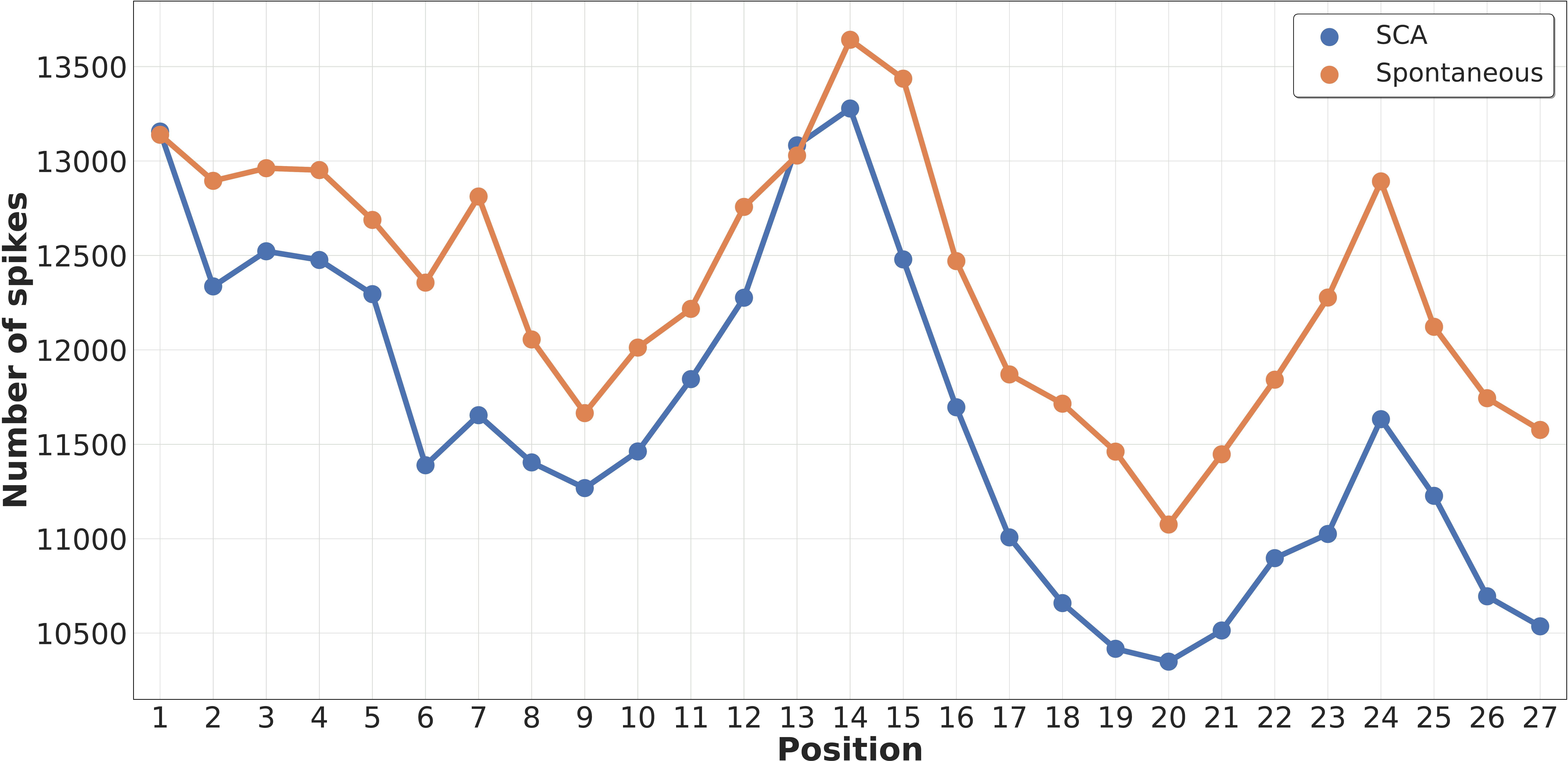}
\end{center}
\caption{Total number of spikes for all neurons of the topology, per position of the optimal path.}
\label{fig:SCA_total_spikes}
\end{figure}

\figurename~\ref{fig:SCA_total_spikes} compares the number of spikes per position of the optimal path between the spontaneous neuronal signaling and an SCA cyberattack. In particular, the SCA cyberattack establishes a value of $40mV$ from the $\mathbb{VI}$ set and defines an aggregation of all three layers of the neuronal topology. We can appreciate the same trend observed in \figurename~\ref{fig:visible_neurons} for the intervening neurons from each of the studied positions. The most prominent peaks are, as previously documented for FLO cyberattacks, delayed one position due to the time required to generate an impact over the neurons. These results can be explained based on the sequential behavior of an SCA cyberattack since the number of attacked neurons raises along time. In addition, this progressive reduction in the number of spikes caused by the attack aligns with the results that will be presented in Section~\ref{subsubsec:SCA_dispersion} for the dispersion metric.

\begin{figure}[h]
\begin{center}
\includegraphics[width=\columnwidth]{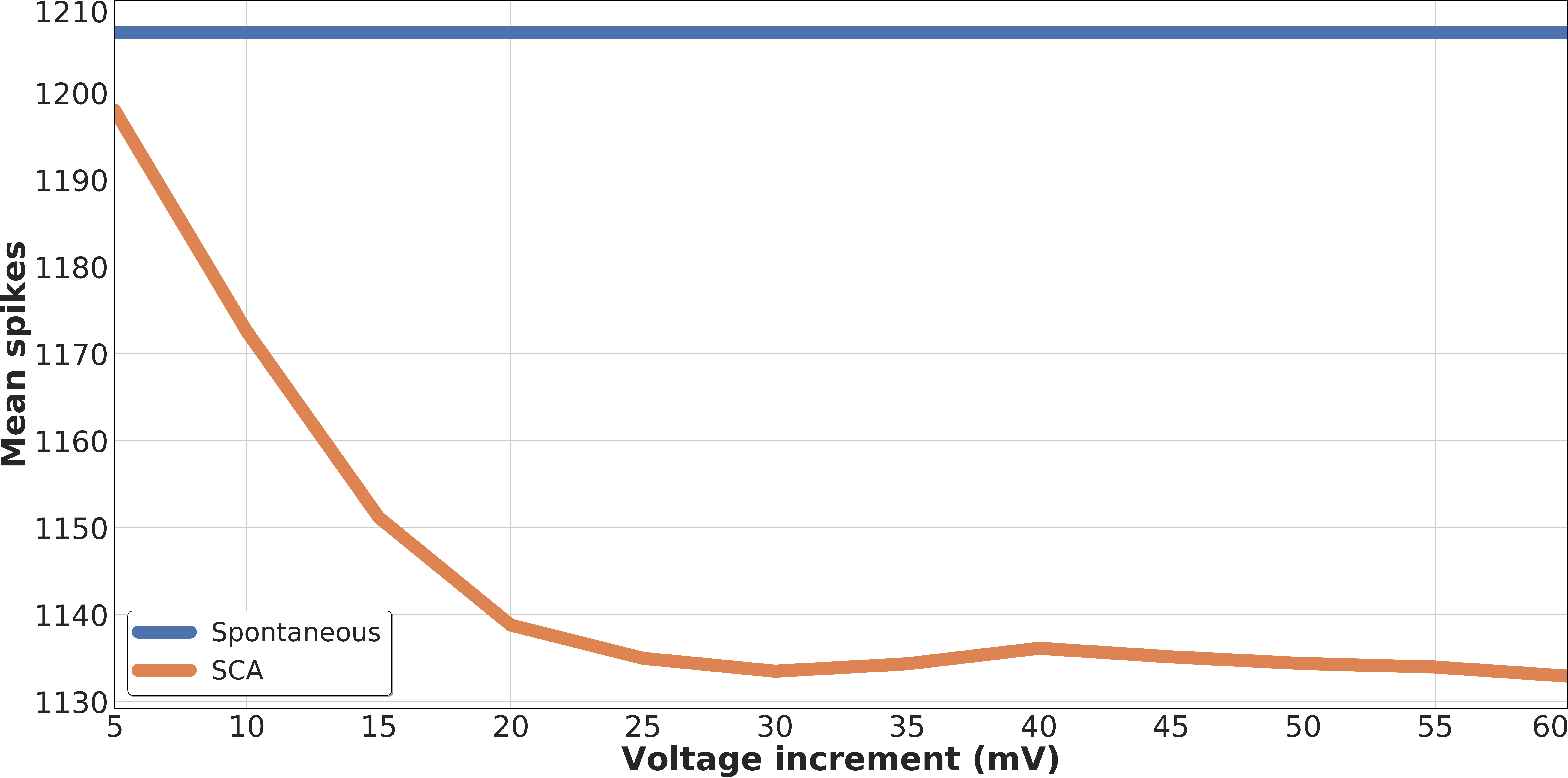}
\end{center}
\caption{Evolution of the spikes mean with different number of attacked neurons and voltage raises, for an aggregation of all positions of the optimal path.}
\label{fig:SCA_global_spikes}
\end{figure}

After this analysis, we evaluated in \figurename~\ref{fig:SCA_global_spikes} the mean of the spikes for the different voltage increases defined in $\mathbb{VI}$, for an aggregation of the three layers of the topology and the positions of the optimal path. We can appreciate that increasing the voltage used to overstimulate the neurons produces a reduction in the number of spikes. It should be noticed that rises higher than $20mV$ do not significantly influence the impact of the attack. Performing an SCA cyberattack with a voltage of $60mV$, the most damaging situation considered, reaches the highest difference in the number of spikes, around 70 spikes compared to the spontaneous behavior.

\begin{figure}[h]
\begin{center}
\includegraphics[width=\columnwidth]{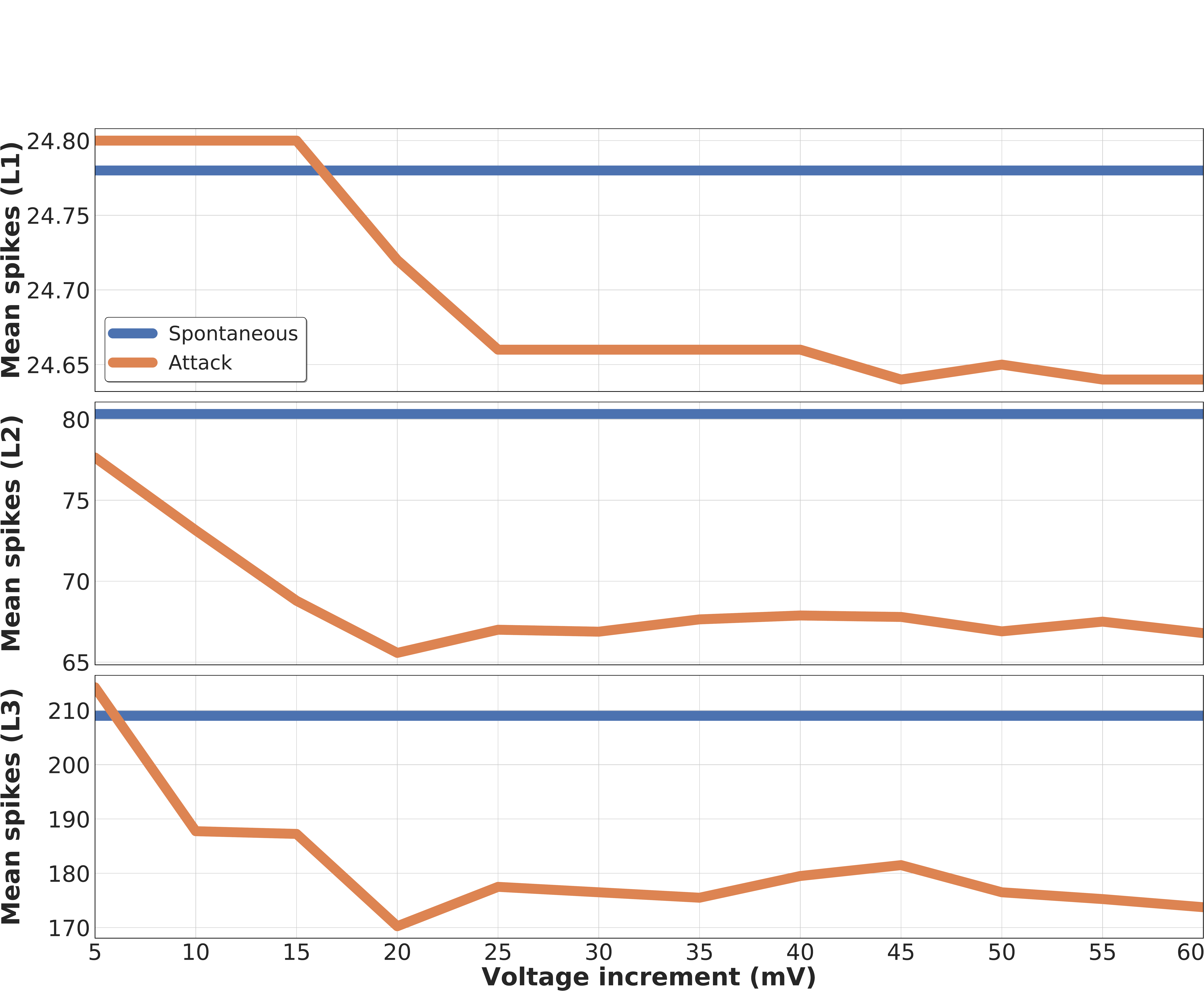}
\end{center}
\caption{Spikes mean for each layer of the topology, focusing on the last position of the optimal path.}
\label{fig:SCA_layers_spikes}
\end{figure}

\figurename~\ref{fig:SCA_layers_spikes} presents a differentiation per layer of the topology for the last position of the optimal path. We can appreciate that, in the first layer, the variation in the number of spikes between different voltage increases is negligible, being  in all cases 24 spikes. Until $15mV$, it presents a small growth of spikes compared to the spontaneous signaling, which benefits of the anticipation of the spikes in time. In more aggressive voltages, the number of spikes gets more reduced than the spontaneous behavior. Moving to the second layer, these differences become more significant, with a number of spikes ranging between 2 and 14 spikes according to the voltage used. This layer presents a general descending trend, reaching the most damaging peak with $20mV$. This trend is common to the third layer, although the range in the number of voltages becomes broader, with a higher difference of 40 spikes compared to the spontaneous signaling. It is interesting to highlight the proliferation of spikes in the third layer when using $5mV$, based on the slight anticipation of spikes in time from the previous layers. 

Comparing these results to those presented in \figurename~\ref{fig:SCA_global_spikes}, we can appreciate in the latter specific differences in the evolution of the impact. In this figure, the most damaging voltage is $60mV$, compared to the $20mV$ highlighted for the second and third layers presented in \figurename~\ref{fig:SCA_layers_spikes}. This situation is explained by the fact that the analysis focused on differentiating the layers only considers one position and, because of that, some minor differences can arise.  

In conclusion, the previous results indicate that performing an SCA cyberattack generates a reduction in the number of spikes, aggravated when the mouse moves across the maze. Increasing the voltage used to overstimulate the neurons does not produce a significant impact with voltages higher than $20mV$. Finally, the number of intervening neurons from each position of the optimal path influences this metric. 

\subsubsection{Percentage of shifts metric}
\label{subsubsec:SCA_shifts}

\begin{figure}[h]
\begin{center}
\includegraphics[width=\columnwidth]{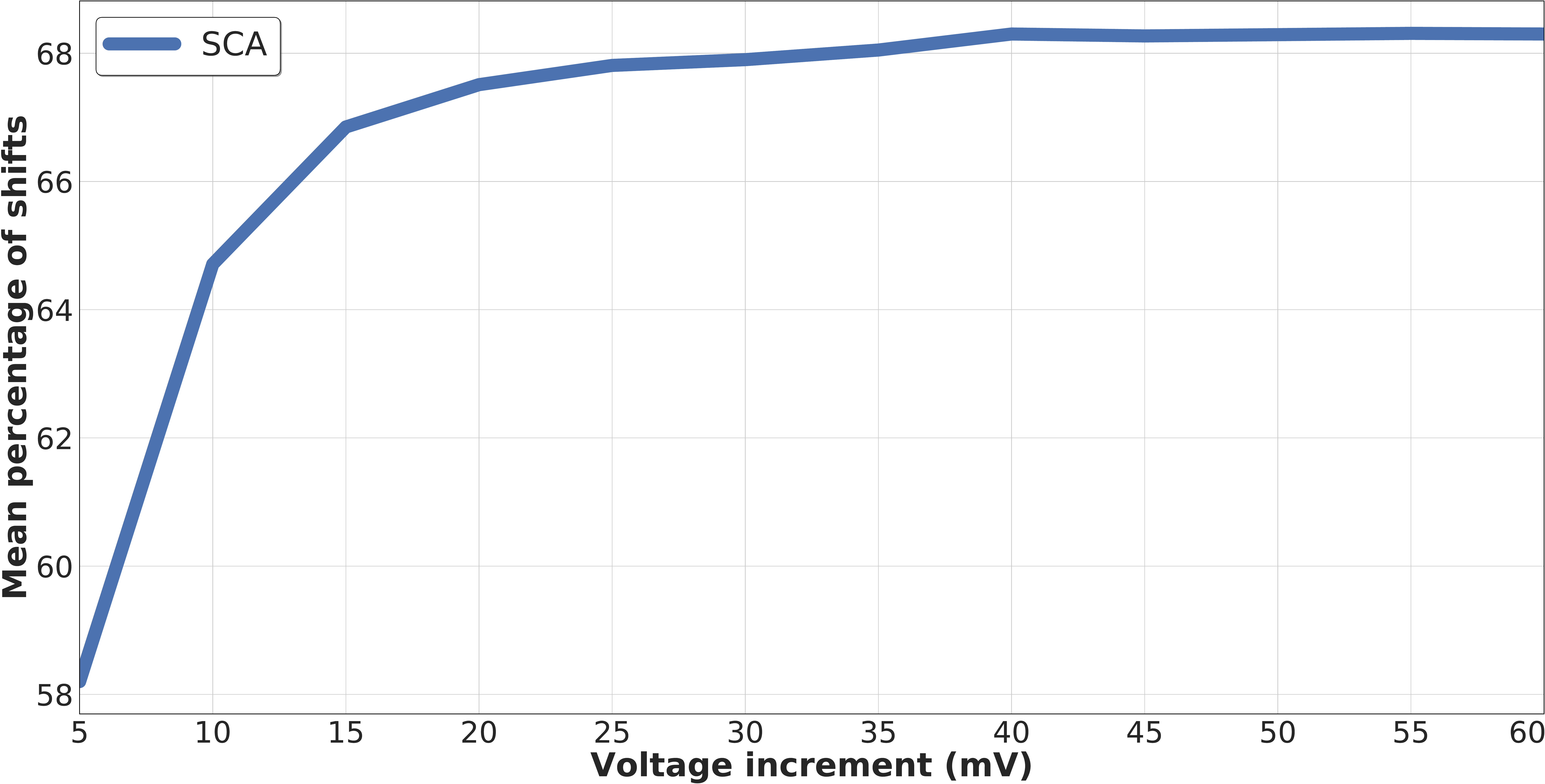}
\end{center}
\caption{Shift percentage mean for an aggregation of all topological layers, aggregating all positions of the optimal path.}
\label{fig:SCA_global_shifts}
\end{figure}

\figurename~\ref{fig:SCA_global_shifts} first presents the results concerning the percentage of shifts for different voltage raises. These results represent an aggregation of the three layers and all the positions of the optimal path. In particular, this figure indicates that the percentage of shifts increases when we raise the voltage used to attack the neurons. We can see that an overstimulation of $5mV$ generates an approximate 58\% of shifts. Slightly increasing this voltage generates considerable impacts, between the range of $5mV$ and $20mV$, reaching a close percentage of 68\%. Finally, increasing the stimulation with voltages higher than $20mV$ does not significantly enlarge the percentage of shifts. These thresholds align with those presented in \figurename~\ref{fig:SCA_global_spikes} for the aggregated number of spikes. 

\begin{figure}[h]
\begin{center}
\includegraphics[width=\columnwidth]{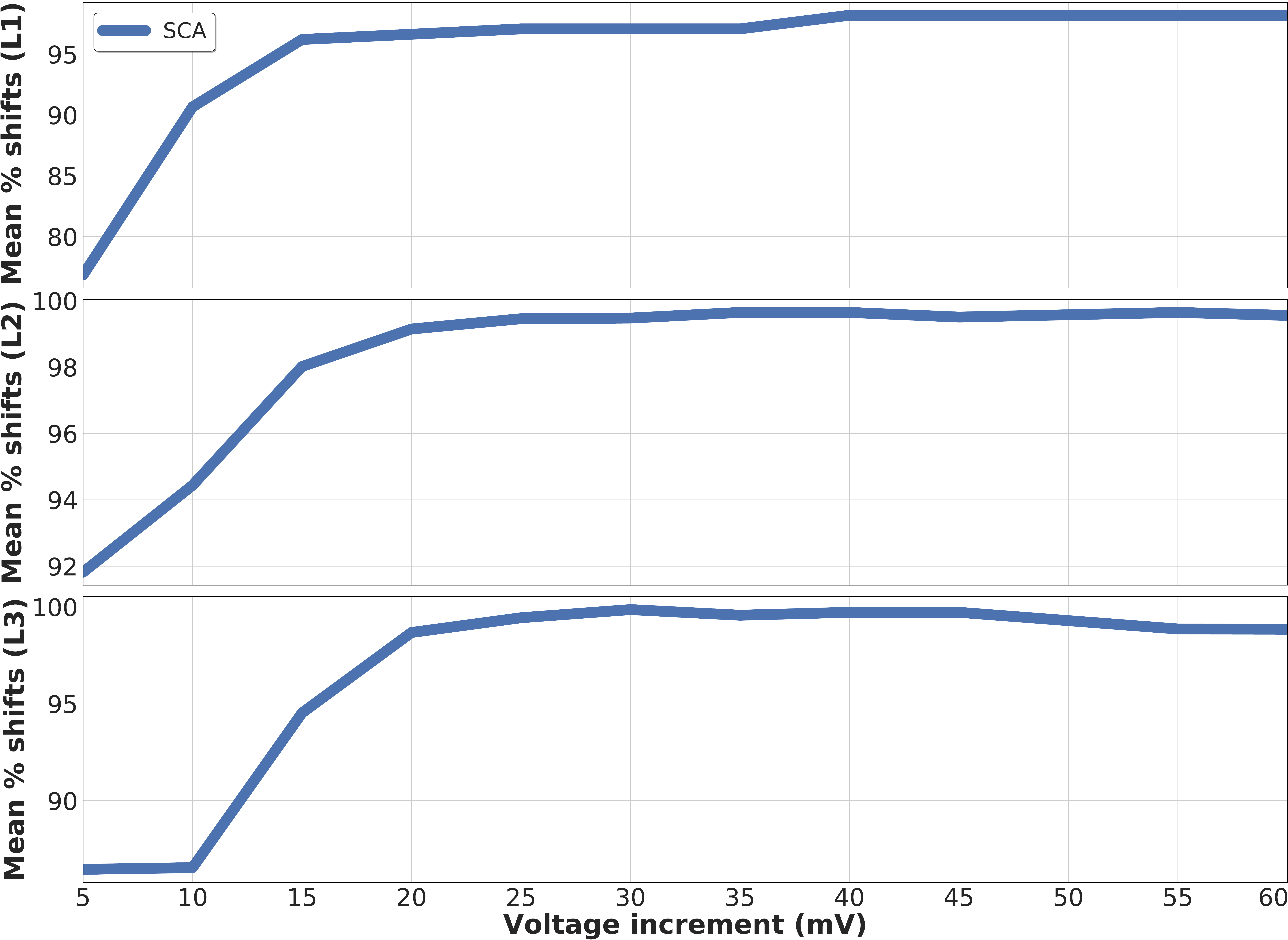}
\end{center}
\caption{Shift percentage mean for each layer of the topology, only focusing on last position of the optimal path.}
\label{fig:SCA_layers_shifts}
\end{figure}

To further explore this metric, we have represented in \figurename~\ref{fig:SCA_layers_shifts}, a differentiation of each layer of the topology for just the last position of the optimal path. We can observe that the range of shifts is lower in the first layer compared to deeper layers, based on the influence that the first layer has on the latter due to the transmitted action potentials. Besides, the growth trend existing in the first layer is more prominent, being similar to the one shown in \figurename~\ref{fig:SCA_global_shifts} for the aggregated analysis of shifts. When we go deeper into the number of layers, we can see that the growth trend is not that aggressive using low voltages, which indicates that the attack progressively loses its effectiveness. It is important to highlight that the ranges shown in \figurename~\ref{fig:SCA_global_shifts} for the percentage of shifts are much higher than those presented in \figurename~\ref{fig:SCA_layers_shifts}. To understand this situation, it is worthy of reflecting on the behavior of SCA cyberattacks. In the first positions of the optimal path, only specific neurons are attacked. When the attack progresses along time, the number of neurons affected by the attack continues increasing. Based on that situation, this last figure focused on the layers presents higher ranges, since they correspond to the last position of the optimal path and, thus, all 200 neurons of the first layer have been affected. 

In conclusion, performing an SCA cyberattack generates a raise in the percentage of shifts. This impact becomes more damaging when the mouse moves across the maze since the number of attacked neurons is more abundant along time. Besides, we can observe a degradation of the impact of the attack in deeper layers, where higher voltages are needed to cause a similar impact in terms of shifts.

\subsubsection{Dispersion metric}
\label{subsubsec:SCA_dispersion}

\begin{figure}[h]
\begin{center}
\includegraphics[width=\columnwidth]{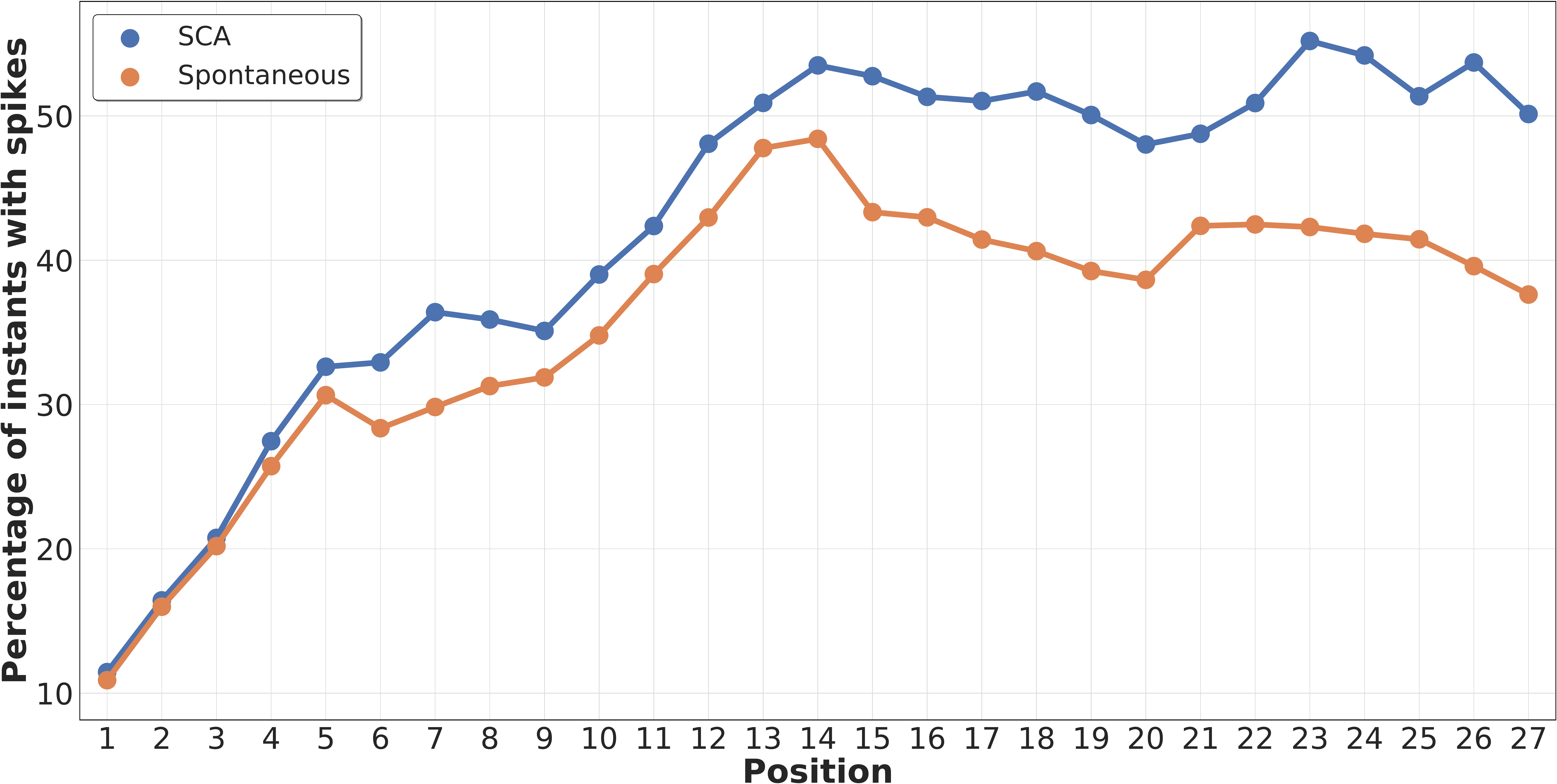}
\end{center}
\caption{Spike dispersion over time for each position of the optimal path.}
\label{fig:SCA_horizontal_dispersion}
\end{figure}

Focusing on the temporal dispersion caused by an SCA cyberattack, \figurename~\ref{fig:SCA_horizontal_dispersion} presents its analysis for each position of the optimal path and the aggregation of all the neurons of the topology. We can observe that performing an SCA cyberattack progressively augments the temporal dispersion, based on the incremental number of attacked neurons over time. In particular, this dispersion is not significant in the first five positions of the optimal path, due to the number of attacked neurons until that moment and the specific connections of our topology. 

\begin{figure*}[h]
\begin{center}
\includegraphics[width=\textwidth]{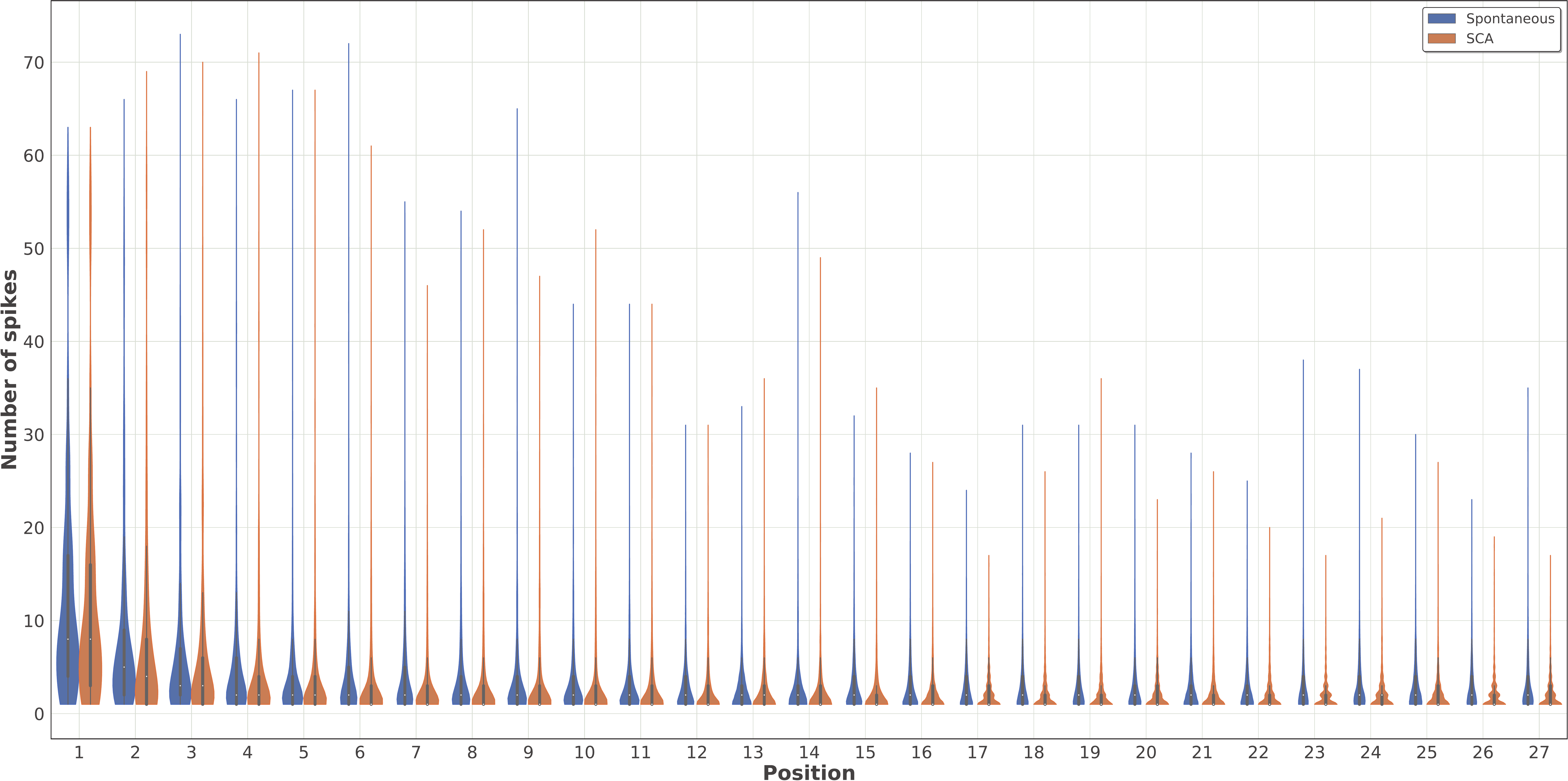}
\end{center}
\caption{Spike dispersion over the number of spikes for each position of the optimal path.}
\label{fig:SCA_vertical_dispersion}
\end{figure*}

After that, we analyze in \figurename~\ref{fig:SCA_vertical_dispersion} the dispersion from the perspective of the number of spikes. In particular, we represent, for each position of the optimal path, a violin distribution of how the spikes behave. We can observe that, in the first five positions, there are no significant visual differences in the distributions, although the median of the distribution start to slightly decrease. This is justified by the reduced number of neurons affected by the attack until that instant. After that position, the differences with the spontaneous behavior progressively augment, both in the peaks in the number of spikes and the shape of the violins. Focusing on the number of spikes, the maximum number of simultaneous spikes presents a reduction, particularly in the last positions. The shape of the violins progressively changes, due to a reduction in their variance, where the number of spikes concentrates at the value of one only spike. That is to say, the majority of the instants in the last positions had only one spike. These results are aligned to those presented in \figurename~\ref{fig:SCA_horizontal_dispersion} for the analysis of the temporal spike dispersion, since both figures indicate that this dispersion increases when the mouse progresses in the maze.

In summary, this metric indicates that performing an SCA cyberattack disrupts the normal neuronal spiking frequency, inducing dispersion in both temporal and number of spikes dimensions. These differences aggravate when the mouse progresses in the maze, based on the sequential functioning of SCA cyberattacks. 

% Comparative between all 3 metrics
The previous three metrics highlight how SCA cyberattacks can affect the spontaneous neuronal activity on our particular topology. We should consider them as different perspectives to analyze a common issue. As previously indicated, an SCA cyberattack progressively induced a decrease in the number of spikes over time, aggravated in deeper layers of the topology. This decrease is strongly related to both dispersion metrics. The attack generates an alteration in the frequency of spikes in time, producing more instants with spikes in the simulation. Specifically, the previous results indicate that in the last positions of the maze, most of the instant only have one spike, which generates a clear difference with the spontaneous activity. The dispersion metric is strongly related to the percentage of shifts since this dispersion will cause a displacement of the spikes in time. In terms of shifts, the attack gets attenuated in deeper layers.

\subsection{Impact comparative between Neuronal Flooding and Scanning}

This last section compares the results previously discussed for FLO and SCA cyberattacks. Focusing on the total number of spikes (\figurename~\ref{fig:FLO_total_spikes} and \figurename~\ref{fig:SCA_total_spikes}), we can observe that an SCA cyberattack generates a more impacting reduction in the number of spikes than the most aggressive FLO configuration. The last positions particularly highlight these differences. 

When we analyze the number of spikes aggregating both positions and layers (\figurename~\ref{fig:FLO_global_spikes} and \figurename~\ref{fig:SCA_global_spikes}), we can appreciate one of the main differences between the attacks. In FLO cyberattacks, we can define as parameters of the attack the number of neurons and the voltage used to attack those neurons. In SCA cyberattacks, we can only specify the voltage, since our implementation affects all neurons of the first layer. Based on that, there is not an immediate comparison between these figures in terms of their trend. Nevertheless, we can compare the most aggressive configuration for each attack to determine which produces the highest reduction of spikes. We can see that SCA presents a slightly higher impact than FLO.

Focusing on the distribution of spikes per layer (\figurename~\ref{fig:FLO_layers_spikes} and \figurename~\ref{fig:SCA_layers_spikes}), we can observe that there are no significant changes between the attacks. In the second one, SCA presents a slightly lower number of spikes. Finally, the third layer amplifies these differences, where SCA has a more significant reduction of spikes. 

In terms of the percentage of shifts (\figurename~\ref{fig:FLO_global_shifts} and \figurename~\ref{fig:SCA_global_shifts}), FLO presents a higher impact on this metric. Extending this comparison for each layer of the topology (\figurename~\ref{fig:FLO_layers_shifts} and \figurename~\ref{fig:SCA_layers_shifts}), we can see that the main difference lies in the first layer, where SCA duplicates its impact since subsequent layers present similar results. Based on that, we can conclude that FLO presents a higher impact on this metric, although the difference in percentages is slight.

There is a clear difference between both attacks in terms of the temporal dispersion metric (\figurename~\ref{fig:FLO_horizontal_dispersion} and \figurename~\ref{fig:SCA_horizontal_dispersion}). FLO has a higher dispersion in the first five positions of the optimal path since the targeted neurons neurons are all attacked in the same instant. After that, SCA evolves in a more damaging way. Focusing on the dispersion based on the number of spikes (\figurename~\ref{fig:FLO_vertical_dispersion} and \figurename~\ref{fig:SCA_vertical_dispersion}), we can observe that FLO is more effective in the first positions. 

This comparative highlights that the inner mechanisms of each attack generates different behaviors in the neuronal activity. FLO is adequate for attacks aiming to disrupt the neuronal activity in a short period of time, affecting multiple neurons in the same instant of time. On the contrary, SCA is a more effective attack for long-term effects, requiring a certain amount of time to reach a significant impact on the neurons. From that threshold, the impact caused on the neurons is more concerning.

%% file: tex/7conclusion.tex
\section{Conclusion}
\label{sec:conclusion}

This work first presents security vulnerabilities of micron-scale BCI to cyberattacks, particularly for implants that can do single-cell or small population sensing and stimulation. Taking these vulnerabilities as a starting point, we describe two novel neural cyberattacks focused on the alteration of neuronal signaling. In particular, we investigated the Neuronal Flooding (FLO) and Neuronal Scanning (SCA), inspired by well-known approaches found in the cybersecurity field. Our investigation is based on a case study of a mouse that learns its navigation within a maze trained by a Convolutional Neural Network (CNN). The CNN was converted into a biological neuronal simulation model representing the workings and functions of real neurons within the brain. The two attacks were applied to the mouse as it migrated through the maze. To evaluate the impact of these attacks on neuronal activity, we proposed three metrics: number of spikes, percentage of shifts, and dispersion of spikes, both over time and number of spikes. 

A number of experiments have demonstrated that both attacks can alter the spontaneous neuronal signaling, where the behavior of these attacks generates distinct differences. FLO attacks all targeted neurons in the same instant of time, while SCA presents an incremental behavior, which requires more time to affect the neuronal activity. Focusing on the results, SCA presents a more damaging impact in terms of the number of spikes, which generates a higher reduction than FLO. In terms of shifts, FLO causes more spikes to differ in time than SCA, although these differences are not very significant. Finally, SCA presents a higher impact on the dispersion of the neurons, both in time and number of spikes. These results are highly dependent on the topology used, the neuronal model utilized to represent the neurons, and the types of neurons used (pyramidal from the primary visual cortex). Because of that, this work should be considered as a first step in the study of cyberattacks affecting spontaneous neuronal signaling.

As future work, we plan to define a taxonomy of neuronal cyberattacks affecting not only overstimulation but also neuronal activity inhibition. We aim to explore how neural cyberattacks can affect realistic neuronal tissues and, in particular, various neural circuits within the cortex. Our research lays the groundwork for security countermeasures to also be integrated into BCI systems that utilize miniature implants for small neuronal population stimulation that can have a tremendous effect on the brain.